\documentclass[twocolumn]{aastex7}

\newcommand{\enzo}{\textsc{Enzo}}

\newcommand{\gcmthree}{g cm$^{-3}$}
\newcommand{\rpunits}{g cm$^{-1}$ s$^{-2}$}

\newcommand{\msun}{M$_\odot$}

\newcommand{\msunpctwo}{M$_\odot$ pc$^{-2}$}

\newcommand{\vtail}{$v_\text{proj}$}
\newcommand{\dtail}{$d_\text{proj}$}

\newcommand{\kms}{km s$^{-1}$}

\newcommand{\tcold}{(T $< 100$  K)}

\graphicspath{{./}{figures/}}

\usepackage{amsmath}
\usepackage{amssymb}
\usepackage{gensymb}
\usepackage{hyperref}
\usepackage{graphicx}
\usepackage{derivative}

\shorttitle{ISM Fallback from Ram Pressure}
\shortauthors{Souchereau et al.}

\begin{document}

\title{Up, Up, and Away? Quantifying ISM Fallback using Ram Pressure Stripping Simulations}

\correspondingauthor{Harrison J. Souchereau}
\email{harrison.souchereau@yale.edu}

\author[0000-0001-5079-1865]{Harrison J. Souchereau}
\affil{Center for Computational Astrophysics, Flatiron Institute, New York, NY 10010, USA}
\affil{Department of Astronomy, Yale University, New Haven, CT 06511, USA}
\email{harrison.souchereau@yale.edu}

\author[0000-0002-8710-9206]{Stephanie Tonnesen}
\affil{Center for Computational Astrophysics, Flatiron Institute, New York, NY 10010, USA}
\email{stephanie.tonnesen@flatironinstitute.org}

\author[0000-0002-9001-6713]{Jingyao Zhu}
\affil{Department of Astronomy, Columbia University, New York, NY 10027, USA}
\email{jingyao.zhu@columbia.edu}

\author[0000-0003-0586-6754]{Jeffrey D. P. Kenney}
\affil{Department of Astronomy, Yale University, New Haven, CT 06511, USA}
\email{jeff.kenney@yale.edu}

\begin{abstract}
    The evolution of the cold interstellar medium (ISM) in satellite galaxies orbiting through massive hosts is an important factor in how they evolve while experiencing ram pressure stripping (RPS), as cold molecular gas clouds are the most difficult ISM component to fully strip and serve as the sites of star formation. 
    We investigate ISM evolution using a suite of hydrodynamical wind tunnel simulations with an intermediate mass ($M_* = 10^{9.7}$ M$_\odot$) galaxy orbiting in a Coma cluster-like environment, varying the disk-wind angle.
    Even if the ultimate fate of a ram pressure stripped galaxy is complete gas removal, we find that cold gas evolves through cycles of outflow and inflow (fallback). 
    We show that fallback can be identified at a wide range of wind angles, but is elevated for angles closer to edge-on and occurs predominantly in a specific quadrant (trailing side, rotating into the wind).
    Most inflow occurs in gas that never leaves an ``inner tail" region that extends to $\sim20$ kpc.    
    We discuss possible reasons for when and why fallback occurs using simple idealized simulations.
    For a highly inclined disk, offset rotational motion is a major driver of fallback, while disk shadowing and cloud growth can act at all wind angles.
    Lastly, we discuss the relative importance of each mechanism at different stages of a galaxy's evolution under ram pressure, and compare our findings with instances of ISM fallback detected in observed RPS galaxies.
\end{abstract}


\section{Introduction}
\label{sec:introduction}

While infalling through a galaxy cluster, the interstellar medium (ISM) of a galaxy interacts with the surrounding intracluster medium (ICM) \citep{gunnInfallMatterClusters1972}. Removal of the ISM through a process known as ram pressure stripping (RPS) involves the direct acceleration of the ISM by the surrounding host halo through which the galaxy orbits \citep[For a recent literature review, see][]{boselliRamPressureStripping2022}.
When ram pressure exceeds the galaxy's gravitational restoring force, this process can completely remove the ISM from the galaxy. 
Because acceleration from ram pressure is inversely proportional to surface density, diffuse gas is rapidly  stripped while nearby dense clouds can survive in the disk \citep[e.g.][]{kenneyEffectsEnvironmentMolecular1989, quilisGoneWindOrigin2000, abramsonHUBBLESPACEESCOPE2014, boselliColdGasProperties2014, leeEffectRamPressure2017, cramerALMAEvidenceRam2020}. This process is believed to be an effective quenching mechanism for spiral galaxies in dense environments, such as galaxy clusters \citep{bookROLERAMPRESSURE2010, bekkiGalacticStarFormation2014, boselliSpectacularTailsIonized2016}. 
However, this process occurs for any satellite galaxies that are orbiting a more massive host, including galaxy groups or the CGM of spiral galaxies \citep[e.g.][]{steyrleithnerEffectRampressureStripping2020, corteseDawesReview92021}.

The study of how cold gas evolves under ram pressure is especially important, as molecular gas clouds are nurseries for star formation in galaxies. Therefore, a thorough understanding of star formation quenching in RPS galaxies requires an understanding of the molecular gas evolution. 
Cold molecular gas is the densest component of the ISM and therefore will be the hardest for ram pressure to strip.
Ram pressure has been observed to efficiently remove lower density gas, but may decouple the lower density ISM from compact molecular clouds that remain in the disk \citep{crowlDenseCloudAblation2005, abramsonHUBBLESPACEESCOPE2014, abramsonHSTIMAGINGDUST2016, cramerALMAEvidenceRam2020}.
Likewise, wind-tunnel simulations with a multiphase ISM such as \cite{tonnesenGASSTRIPPINGSIMULATED2009} show that high density gas is stripped less than low density gas.
Despite this, molecular gas can be stripped, and has been detected in the tails of RPS galaxies \citep[e.g.][]{boselliColdGasProperties2014, jachymMolecularGasDominated2017, 
morettiGASPXXIIMolecular2020, poggiantiGASPXXIIIJellyfish2019}.
Evidence for direct molecular gas stripping has been seen in some galaxies \citep[e.g][]{cramerALMAEvidenceRam2020}.

However, the gas surface density of the ISM is not the only factor affecting the efficiency of gas removal by RPS.
Because disk galaxies are rotating, the effects of ram pressure can change depending on the the angle between the galaxy disk normal vector and the wind vector, also known as the \textit{disk-wind angle}.
Infalling galaxies with a random initial distribution of orientations are more likely to have a velocity vector closer to the disk plane than the disk normal vector. 
Moreover, given that the wind angle will change for almost all orbits, it is expected that most galaxies should experience a highly-inclined (closer to edge-on) phase at some point in their orbit.
 It is now well-established that a more inclined wind leads to less efficient stripping \citep{roedigerRamPressureStripping2006, jachymRamPressureStripping2009, leeDualEffectsRam2020, akermanHowRamPressure2023, sparreMagnetizedThermallyUnstable2024}. 

When considering a highly-inclined disk-wind angle, the rotation of the disk now factors into the ram pressure strength, leading to asymmetries across the disk.
The different sides should be affected differently due to how the ram pressure wind applies torque into the rotating ISM gas. On the side of the galaxy rotating with the wind, the ram pressure torque acts to increase angular momentum, and the gas is driven outwards, leading to preferential stripping on this side \citep[e.g.][]{souchereauALMAJELLYHighResolution2025}. On the side rotating into the wind, torques decrease angular momentum, and gas will be driven radially inwards \citep[e.g.][]{akermanHowRamPressure2023}.
Furthermore, the angle between the gravitational restoring force and the ram pressure wind vector is constant for a face-on wind, but varies when the wind is inclined.

The fate of ISM gas that has been pushed out of the disk, but not accelerated sufficiently to escape the galaxy's gravitational potential well, is important for understanding how the ISM (and consequently, galaxy stellar components) can reassemble before a galaxy is fully stripped.
Most of the perturbed (but not fully stripped) ISM will be expected to reside close to the galaxy body it was stripped from.
Under certain circumstances, this material can return to the disk; a process called ``fallback" which has been observed in some RPS galaxies (NGC 4921: \cite{cramerMolecularGasFilaments2021}, NGC 4858: \cite{souchereauALMAJELLYHighResolution2025}). 
Fallback has also been identified in other hydrodynamical simulations of ram pressure \citep{schulzMultistageThreeDimensional2001, vollmerRamPressureStripping2001, tonnesenTAILSTRIPPEDGAS2010, leeDualEffectsRam2020, choiRamPressureStripping2022, akermanHowRamPressure2023, zhuWhenHowRam2023, sparreMagnetizedThermallyUnstable2024}.

Although not typically addressed in the same context, gas removed from the galaxy disk, but not yet fully stripped, can be viewed as a component of the circumgalactic medium (CGM) of satellite galaxies. Unlike the typical picture in the baryon cycle of internal processes driving outflows of gas from the ISM to CGM \citep[e.g.][]{wrightBaryonCycleModern2024}, a ram pressure wind can also be a driver of gas outflows. Hydrodynamical wind-tunnel simulations have shown that the CGM of dwarf galaxies is almost immediately stripped by ram pressure, removing the capacity for gaseous inflows via the ``typical" channels present in the CGM \citep{zhuItsBreezeCircumgalactic2024}. Therefore, fallback due to ram pressure may be one of the only ways to achieve gaseous inflows for satellite galaxies orbiting within dense environments.

Several open questions about fallback and it's role in galaxy evolution remain unanswered: How much fallback is to be expected, and from what phase of gas? Can fallback be localized to specific regions of a stripped gas tail? During what parts of a satellite's orbit does fallback occur? Answering these questions is important to fully understand how galaxies evolve under the effect of ram pressure.
In this work, we will examine how the asymmetric impact of an inclined wind on a rotating satellite affects how gas is removed, and can eventually return to, its disk.

The paper is outlined as follows. In Section \ref{sec:sim-setup} we briefly describe our simulation setup and initial conditions, as well as how we initialized our satellite galaxy and the ram pressure wind. In Section \ref{sec:global-properties} we show how the ISM disk evolves for each simulation run, and measure the asymmetrical effect of ram pressure across the disk using quadrant analysis. 
In Section \ref{sec:measuring-fallback}, we highlight two separate methods of measuring fallback, and then apply them to our simulations in order to identify where fallback occurs in the tail, and how much inflowing ISM is present. In Section \ref{sec:why-fallback-occurs}, we examine the possible mechanisms generating fallback, and in Section \ref{sec:importance-and-comparisons} we discuss the relative importance of each mechanism and compare to observed cases of RPS galaxies with fallback. 
Finally, we summarize our findings in Section \ref{sec:conclusion}.

\section{Simulation Setup}
\label{sec:sim-setup}

We run a suite of high resolution, 3-dimensional ``wind tunnel" simulations using the adaptive mesh refinement (AMR) code \enzo{} \citep{bryanENZOAdaptiveMesh2014}. The \enzo{} simulation initial parameters are the same as in \cite{zhuWhenHowRam2023}, to which we refer the reader for a more detailed description. We briefly summarize our setup here. 

The simulation volume is 162 kpc per side, with a base-level resolution of $128^3$ grid cells. From this, we allow up to 5 levels of resolution refinement based on grid cell density, with a smallest possible cell width of 40 pc. Comparisons of our results to a lower resolution run can be found in Appendix \ref{app:caveats-resolution}, where we show that the lower resolution does not affect our main conclusions.
Radiative cooling is modeled using the \texttt{Grackle} chemistry and radiative cooling library\footnote{\href{https://grackle.readthedocs.io/en/latest/}{https://grackle.readthedocs.io/en/latest/}} \citep{bryanENZOAdaptiveMesh2014, kimAGORAHighresolutionGalaxy2014}. The star formation recipe, from \cite{goldbaumMASSTRANSPORTTURBULENCE2015}, creates new star particles at a 5\% efficiency in grid cells where the gas density reaches the Jeans criterion and a minimum number density of $10$ cm$^{-3}$. These star particles proceed to act on the gas through stellar and supernova feedback, following the \cite{goldbaumMASSTRANSPORTTURBULENCE2016} feedback model. 

As we will describe in more detail below, the modeled galaxy and ram pressure wind profile have been tuned to generally match the Coma cluster galaxy NGC 4858 and its orbit through the cluster. 
This galaxy, with a stellar mass of $4.9\times10^{9}$ \msun, is a barred spiral jellyfish galaxy currently experiencing strong ram pressure stripping with a line-of-sight velocity of $>2000$ \kms{} relative to the cluster mean. 
As discussed in \cite{souchereauALMAJELLYHighResolution2025}, ALMA observations of the molecular gas in NGC 4858 have revealed the best example of gas fallback to date, making this galaxy  a suitable model on which to base our simulation suite.

\subsection{The Galaxy}
\label{sec:isolated-galaxy}

The galaxy is placed at the center of the simulation volume. It originally consists of static dark matter and stellar gravitational potentials, and a live gas distribution. A summary of the components of the simulated galaxy is in Table \ref{tab:potential}.

\begin{figure*}
    \centering
    \includegraphics[width=\textwidth]{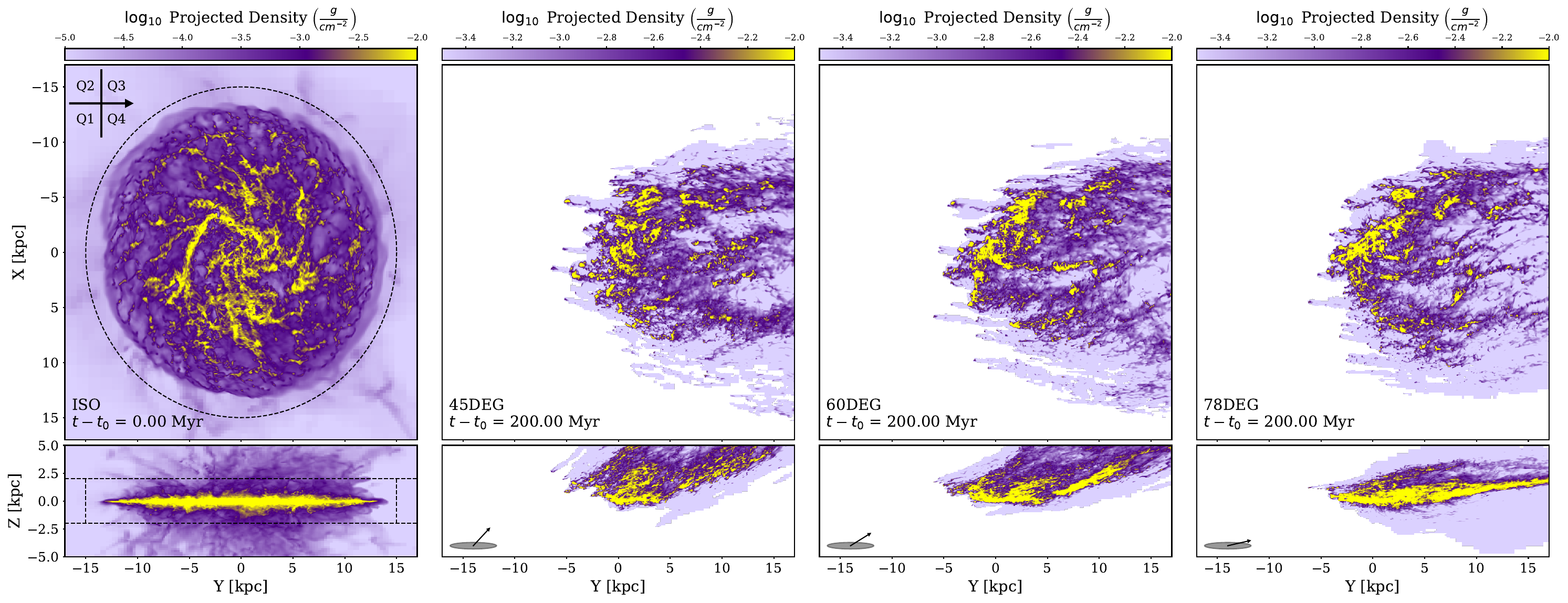}
    \caption{Density projections of the ISM ($Z/Z_\odot > 0.25$) in our simulations. The top row shows Y-X (top down) projections, whereas the bottom row shows Y-Z (edge-on) projections.
    The leftmost column shows the galaxy immediately before wind is injected into the box ($t=691$ Myr, snapshot 158). A diagram showing the quadrant divisions for this galaxy (as well as the wind direction), discussed in Section \ref{sec:quadrants}, is displayed in the top left panel. The dashed lines indicate the boundaries of our \textit{static} disk volume (defined in Section \ref{sec:selection-criteria}).
    The remaining three columns show snapshots of the different wind runs, where the time elapsed is the same (400 Myr) but the disk-wind angle is different.
    We indicate the angle of the wind with the white arrow in each lower panel. }
    \label{fig:galaxy-init}
\end{figure*}

Our static gravitational potential is a composite potential consisting of dark matter and an initial stellar disk.
The dark matter potential is described by the Burkert potential \citep{moriGasStrippingDwarf2000}. This potential is generally used to describe the dark matter potentials of dwarf satellites but can also be used to model the potential in higher-mass galaxies. The static pre-existing stellar disk is modeled using a Miyamoto-Nagai potential \citep{miyamotoThreedimensionalModelsDistribution1975}. 

The gas density is modeled with a softened exponential disk \citep[][see their Equation 1]{tonnesenGASSTRIPPINGSIMULATED2009, tonnesenTAILSTRIPPEDGAS2010}. From this gas density profile we obtain a central density of $1.68\times10^{-24}$ \gcmthree. 
The temperature and pressure of the gas are calculated to maintain hydrostatic equilibrium
(along the direction perpendicular to the disk)
with the surrounding ICM. 
The rotational velocity of the gas is then determined by taking into account both the gravitational potential force and the pressure gradient in the disk. All of the gas in the disk is initially set to a metallicity of $Z/Z_\odot = 0.3$.

\begin{table}[t]
\centering
 \caption{Initial parameters of the cluster galaxy. The Burkert density profile is characterized by the central density $\rho_{d0}$ and the core radius $r_0$. The stellar disk and gas disk are each characterized by a total mass ($M_*, M_g$), a scale radius ($a_*, a_g$), and a scale height ($b_*, b_g$).}
 \footnotesize
 \begin{tabular}{c c c c}
         Component & Parameter &  & \\
         \hline
         Burkert DM Halo & $\rho_{d0}$ & $ 4.77\times10^{-25}$ & g cm$^{-3}$ \\
         & $r_0 $ & $ 16.42$ & kpc \\
         M-N Stellar Disk & M$_* $ & $ 4.89 \times 10^{9}$ & M$_\odot$ \\
         & $a_* $  $b_*$ & $ 2.2$ \hspace{1em} $0.5$ & kpc \\
         Exponential Gas Disk & M$_g $ & $ 4.89 \times 10^{9}$ & M$_\odot$ \\
         & $a_g $ $b_g$ & $ 3.96$ \hspace{1em} $0.5$ & kpc \\
         \hline

    \end{tabular}
    \label{tab:potential}
\end{table}

We allow the galaxy to evolve in isolation for $\sim700$ Myr before injecting the wind into the box. This is a longer relaxation time than in other works such as \cite{tonnesenStarFormationRam2012} ($\sim200$ Myr) and \cite{akermanHowRamPressure2023} ($\sim230$ Myr), in order to achieve a more realistic initial gas disk morphology in addition to stabilized star formation. 
The gas structure of the galaxy can be observed in the left panel of Figure \ref{fig:galaxy-init}, which shows gas density projections (face-on and edge-on in the top and bottom panels, respectively) of the disk immediately before wind is injected into the simulation volume. 
This is our isolated galaxy, from which every wind run diverges. 
After this stage of initial evolution, the disk gas has increased in average metallicity to $Z/Z_\odot = 0.45$.
We continue simulating the galaxy in isolation for as long as our wind runs in order to have a baseline case with which to compare. 

\subsection{Wind Prescription}
\label{sec:wind-setup}

To simulate a plausible ram pressure wind for our galaxy, we first produce a time series of the galaxy's velocity by modeling the galaxy's orbit based on the projected distance and estimated total velocity of NGC 4858 \citep[from][]{souchereauALMAJELLYHighResolution2025}. Based on these initial conditions, the orbit is then generated using the \textsc{Gala} galaxy dynamics package \citep{price-whelanAdrnGalaV1812024}. 
We generate a smooth gravitational potential for the Coma cluster, with a virial mass of $1.2\times10^{15}M_\odot$ and $c=9.3$ \citep{lokasDarkMatterDistribution2003}. We then calculate the orbit on our galaxy both backwards and forwards in time to obtain the satellite's clustercentric radius and velocity over many complete orbits through the model environment.

To determine the ram pressure, we also need to calculate the gas density along the galaxy's orbit. We adopt an ICM gas density profile from \cite{mirakhorCompleteViewOutskirts2020}. We fit a Beta profile \citep{arnaudVmodelIntraclusterMedium2009} to the density profile, which has the form

\begin{equation}
    n(r) = n_0\left[1 + \left(\frac{r}{r_c}\right)\right]^{-3\beta/2} .
\end{equation}

\noindent After least-squares minimization, we obtain parameters of $\beta=0.53$, $n_0=2.26\times10^{-26}$ \gcmthree (R), and $r_c=83$kpc out to $\sim900$ kpc, well beyond the furthest radial extent of our simulated galaxy's orbit. 
We then use the estimated total velocity of the galaxy as well as the density profile to set the wind injected into our box. 

The isothermal (T = $10^7$ K, $Z/Z_\odot = 0.1$) wind is introduced at the box boundary (y=0) after $690$ Myr of isolated evolution.  Because the galaxy is located in the center of the box and the wind is introduced at the boundary, the wind does not actually reach the galaxy for another $\sim50$ Myr. We highlight that the time the wind arrives at the disk is approximately 740 Myr, and the point of peak ram pressure is at 1330 Myr. There is a second peak at $\sim2500$ Myr, but all of the galaxies in the wind runs have been completely stripped before a second pericentric passage.
We often quote the time after the wind reaches the disk ($t-t_0 = t - 740 \text{ Myr}$) instead of the total simulation time.

\subsection{Simulation Suite}

The winds injected into our simulations flow in the $y-z$ direction, with no $x$ component. The total ram pressure strength as a function of time is shown in the top panel of Figure \ref{fig:disk-gas-mass-evol}, estimated using a small volume placed in front of the galaxy at (40, 40, 40) kpc \citep[as in][]{zhuWhenHowRam2023}. The total ram pressure strength is kept consistent between runs. However, each run has a different disk-wind angle, or the  proportions of total velocity that are applied to the $v_z$ and $v_y$ components of the wind. Our disk-wind angles (and run names) are $45\degree$ (45DEG), $60\degree$ (60DEG), and $78\degree$ (78DEG), and remain constant throughout each simulations. 
All of our simulations run until stripping is complete (i.e., no gas belonging to the original, unperturbed galaxy, is remaining within the original extent of the gas disk. See the bounding box in Figure \ref{fig:galaxy-init}). We highlight that all three of our wind simulations ultimately reach a fully stripped state.
Pictorial representations of the three disk-wind angles, as well as the bounding box capturing the extent of the unperturbed gas disk are shown in Figure \ref{fig:galaxy-init}.

\subsection{Selection Criteria and Derived Fields}
\label{sec:selection-criteria}

In this paper we will be analyzing the distribution and velocity of gas that originates from within the stripped galaxy.
To achieve this we define a set of physical criteria.
We use \textsc{yt}: a \textsc{Python}-based analysis toolkit for volumetric data to analyze our simulation suite \citep{turkYtMulticodeAnalysis2011} \footnote{See \href{https://yt-project.org/}{https://yt-project.org/}}. This provides both direct numerical analysis, as well as plotting routines used in this paper.

\subsubsection{Identifying ISM Gas}

We use a metallicity of $Z/Z_\odot \geq 0.25$ to distinguish cells that consist of ISM gas (which has been enriched during the first $\sim700$ Myr of isolated evolution) from metal-poor ICM gas. This is higher than the metallicity of the wind ($Z_\text{wind} = 0.1$), but somewhat less than the initial galaxy disk metallicity ($Z_\text{disk} = 0.3$), allowing for ISM gas to mix slightly with its surroundings and still satisfy this requirement. 
Even with an average disk metallicity of $Z/Z_\odot = 0.45$ at the time the wind reaches the disk, a cell above our metallicity cut has nearly half of its mass from the galaxy. We tested a more conservative ISM threshold of $Z/Z_\odot = 0.35$ on our fallback measurements (see Section \ref{sec:quantifying-fallback}) and found it does not significantly change our results.
Following this, we apply temperature cuts to define cold \tcold{} ISM gas.
We also tested $T < 300$ K and $T < 1000$ K and found qualitatively similar results.

\subsubsection{Defining the Disk Region} \label{sec:disk_definitions}


For our simulations, we wish to separate tail gas from disk gas in order to capture inner tail evolution at all stages of an RPS event, where the size of the gas disk does not remain static with time. We achieve this by measuring a \textit{truncation radius} $R_{\text{trunc}}$, which is a size estimate of the cold gas disk of the galaxy at each simulation snapshot. 
We measure the radius of the gas disk for each snapshot by taking a spherical radial profile of the leading side of galaxy's cold gas distribution.
The leading and trailing sides split the galaxy along a vertical plane ($y < 0$ for leading, and $y > 0$ for trailing), where the leading (trailing) side of the galaxy is the side closest to (furthest from) the wind front. 
We exclude the trailing side as this is likely to be contaminated by the tail. We then define the truncation radius as where the profile (smoothed with a moving-average filter to reduce small scale variations) reaches $10^{-26}$ \gcmthree, which is the same density limit adopted in \cite{roedigerRamPressureStripping2006} and determined within to be an appropriate representative density limit. This is higher than the maximum density reached by the ram pressure wind ($\rho_{\text{ICM, peak}} \approx 2.2\times10^{-27}$\gcmthree).
We find that this choice satisfactorily captures the size of the cold gas disk for almost every snapshot \footnote{In some cases, especially in later stages when the remaining cold ISM quantity is low, the gas can often ``slosh" around the gravitational potential. During this phase, perturbed gas can fall back and punch through the remaining disk. This briefly increases the amount and distribution of cold ISM gas on the leading side, leading to a slight overestimate in the truncation radius.}.

When defining the radial extent of the galaxy ``disk", there are two options, both of which we will use in this work. The first is of measurements inside a cylinder enclosing the initial, unperturbed galaxy that is fixed over time. This is our \textit{static} disk, given by $R < 15$kpc and $|z| < 2$kpc. The boundaries of this selection volume can be seen in both panels of Figure \ref{fig:galaxy-init}. This is slightly larger than the bulk of the ISM disk, which allows for smaller-scale fluctuations due to star formation feedback.  The other disk measurement takes into account the truncation radius of the galaxy disk. We keep the same height requirement ($|z| < 2$kpc) but instead fix the radius to the truncation radius $R_\text{trunc}(t)$. This will allow for a more refined exploration of some inner-tail processes that might remain entirely enclosed within the initial, static disk.

\subsubsection{Quadrants}
\label{sec:quadrants-criteria}

\begin{table}[]
    \centering
    \begin{tabular}{c|c}
        Quadrant & Location  \\
        \hline
        1 & Leading side, rotating into wind \\
        2 & Leading side, rotating with wind \\
        3 & Trailing side, rotating with wind \\
        4 & Trailing side, rotating into wind \\
    \end{tabular}
    \caption{Summary of quadrants and their location with respect to the wind and galaxy rotation.}
    \label{tab:quadrants}
\end{table}

A non-uniform ram pressure strength due to an inclined wind on a rotating galaxy is expected to create asymmetries in the galaxy's ISM distribution. These symmetries are expected to develop between the leading/trailing sides of the galaxy, as well as on the sides of the galaxy rotating with and against the ram pressure wind.
To first order, we can examine the behavior of the gas and how it varies azimuthally by subdividing the galaxy's flux distribution into four \textit{quadrants}. 
For our suite, the quadrant cuts are along the $x$ and $y$ axes of the simulation box. The quadrant order follows the rotation of the galaxy, beginning with Q1 on the leading side, rotating into the wind. We emphasize that this analysis can be easily applied to real galaxies so long as (1) the galaxy rotation and (2) the position angle of the ram pressure wind can be estimated. The quadrants are summarized in Table \ref{tab:quadrants}. A visual representation of the quadrants is also shown in the first panel of Figure \ref{fig:galaxy-init}.

\section{Evolution of the ISM Disk}
\label{sec:global-properties}

Before describing the evolution of gas in the tail, we first look at how the inclination angle of the RP wind affects the evolution of the ISM disk. We first explore the evolution of the size and total mass of the ISM disk, and then subdivide the ISM distribution into quadrants to reveal RP-induced asymmetries.

\subsection{Disk Gas Masses and Radii}
\label{sec:disk-gas-mass}

\begin{figure}
    \centering
    \includegraphics[width=0.5\textwidth]{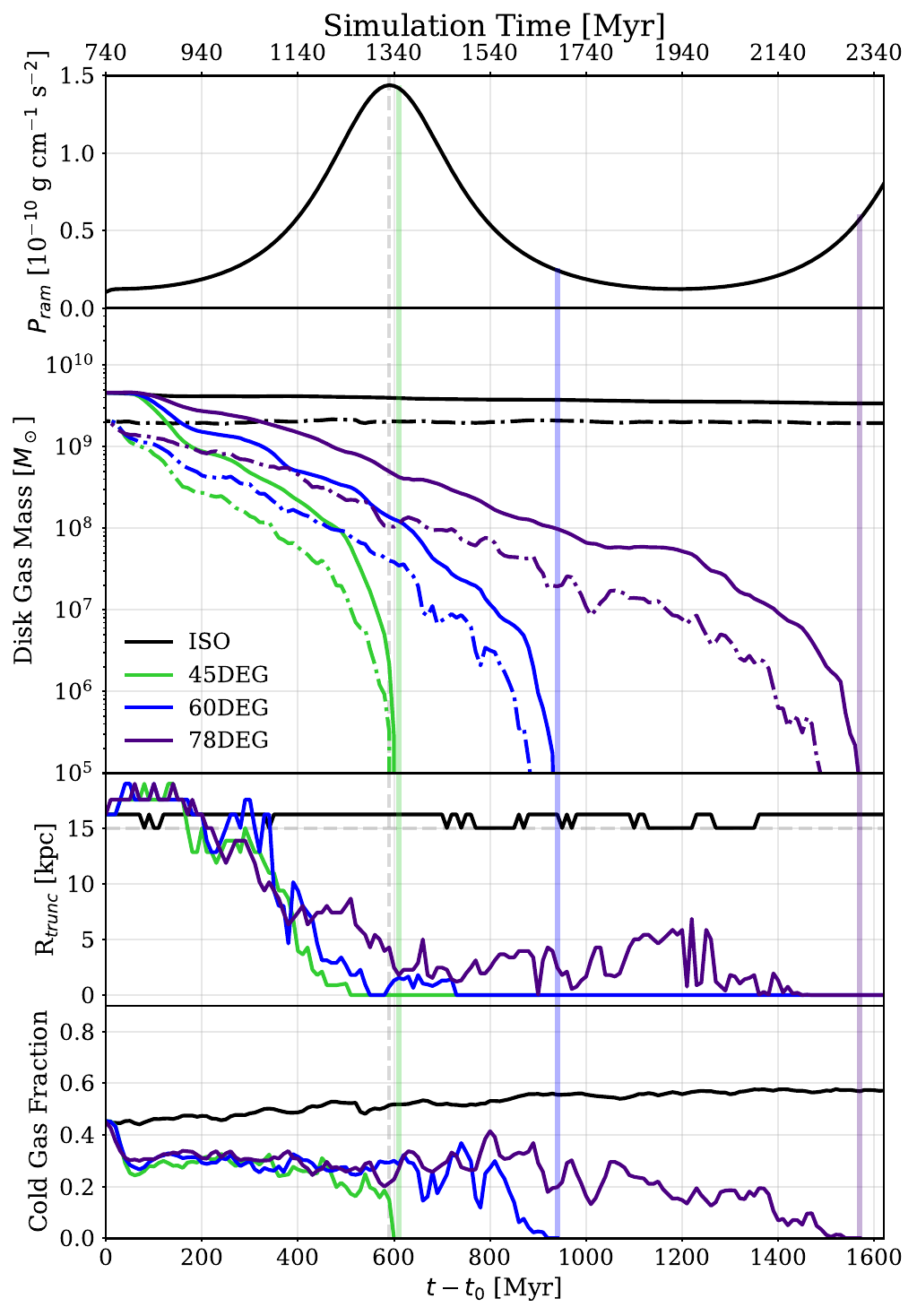}
    \caption{Evolution of the galaxy ISM disk over time. 
    \textbf{Panel 1}: The strength of ram pressure over time, measured using a small spherical volume placed near the disk.  
    \textbf{Panel 2}: Total gas mass (within the static disk) vs time for each simulation. The solid and dashed lines represent the ISM ($Z/Z_\odot > 0.25$) and cold ISM \tcold{} mass profiles, respectively.
    \textbf{Panel 3}: The evolution of the truncation radius (described in Section \ref{sec:selection-criteria}).
    \textbf{Panel 4}: The fraction of ISM disk gas with temperature $T < 100$K.
    The vertical dashed line indicates the point of peak ram pressure at pericentric passage, and the colored lines indicate the approximate time of complete stripping for the three runs. 
    }
    \label{fig:disk-gas-mass-evol}
\end{figure}

To begin our analysis, we first look at the time evolution of global properties in the galaxy's gas disk.
Figure \ref{fig:disk-gas-mass-evol} shows the evolution of the disk's gas mass, disk truncation radius, and disk cold gas fraction over time. We indicate the approximate times when each simulation run reaches a fully stripped state with vertical colored lines. The colors for each run (ISO = black, 45DEG = green, 60DEG = blue, 78DEG = purple) are consistent throughout this work.

The second panel shows the evolution of total gas mass in the static disk (defined in Section \ref{sec:disk_definitions} as having a 15 kpc radius). The solid lines are for ISM gas (metallicity cut only), whereas the dash-dot lines show the evolution of cold ISM gas \tcold. While the ISO run does slowly lose gas mass over time due to star formation and stellar winds blowing out material, it remains mostly constant. 
There is a clear trend with stripping efficiency and disk-wind angle, with an increasingly inclined wind (closer to edge-on) corresponding to a slower stripping rate. 
The most inclined 78DEG run takes almost an additional Gyr to reach a fully stripped stage compared to the 45DEG run.
\footnote{We compute the approximate time of complete stripping by measuring when the ISM gas mass in the static disk drops below $ 10^4$ M$_\odot$ for the first time, which is approximately the mass of a single cell at our star formation threshold density ($n = 10$ cm$^{-3}$)). Given the steep drop of gas mass at later times, our measured times of complete stripping would change very little even if raising or lowering this threshold by an order of magnitude.}
This trend with disk-wind angle is consistent with findings from other hydrodynamical wind tunnel simulations \citep[e.g.][]{roedigerRamPressureStripping2006, jachymRamPressureStripping2009}.

The third panel shows the measured truncation radius $R_{\text{trunc}}(t)$. For all runs we tend to see a consistent decrease up until pericenter, which is indicated by the gray vertical line. The 45DEG run is fully stripped by then and remains at zero, as does the the 60DEG with the exception of a small bump back to 1-2 kpc for approximately 100 Myr after reaching pericenter. The 78DEG, however, never reaches a radius of zero during first infall and actually appears to grow after pericenter, maintaining an approximate size of 3-5 kpc up until $t-t_0 = 1200$Myr before shrinking completely.
This can be attributed to the ram pressure profile, where the 78DEG survives the first pericentric passage and enters a period of reduced ram pressure strength. It is not until beginning the second plunging orbit when the remaining ISM is finally evacuated.
The truncation radius also reaches zero before the gas in the disk is fully stripped. This is due to some gas still remaining on the trailing side of the galaxy (so within the bounds of the static disk), but no gas remains to be measured on the leading side.

While the truncation radius for all three runs decreases at a similar rate, the 78DEG run loses gas much slower than the other two runs. This is likely because the disk gas mass is being measured using the static disk, which has a height of $\pm2$ kpc. For the less inclined winds, more gas is pushed above (and out of) the static disk, meaning those runs will lose disk gas mass rapidly, whereas for the 78DEG run, the gas is being pushed towards the trailing side. This means the truncation radius for the 78DEG run can decrease at a similar rate to the other runs, while the mass decreases much slower.

In the bottom panel we show the evolution of the cold gas fraction. Immediately upon the wind striking the galaxy, there is a reduction by about 30\% in the galaxy. There appears to be a weak dependence on angle, with less inclined winds reducing the gas fraction the most. 
The cold gas fraction never recovers from this initial reduction. After the initial drop, it remains approximately constant for an extend period of time before dropping at later times.
We suspect that the early decline in cold gas fraction is at least in part due to heating of the cool/cold ISM by the hot ICM wind.
A second effect might be due to ICM wind mixing with warmer ISM gas, becoming enriched and thus increasing to the total ISM gas. Given that there aren't noticeable increases in total ISM mass in Figure \ref{fig:disk-gas-mass-evol}, this is likely an insignificant effect.
Some observational studies, such as \citep{brownVERTICOVirgoEnvironment2021}, have shown that the molecular gas fractions of RPS galaxies tends to be regular, which disagrees with our measured drop in cold gas fraction.
However, we are only marginally resolving gas at temperatures representative of molecular gas ($\leq 10$ K), so we cannot easily compare our simulations to molecular gas observations directly.

\subsection{Quadrant Analysis: Gas Loss Asymmetry}
\label{sec:quadrants}

\begin{figure*}
    \centering
    \includegraphics[width=\textwidth]{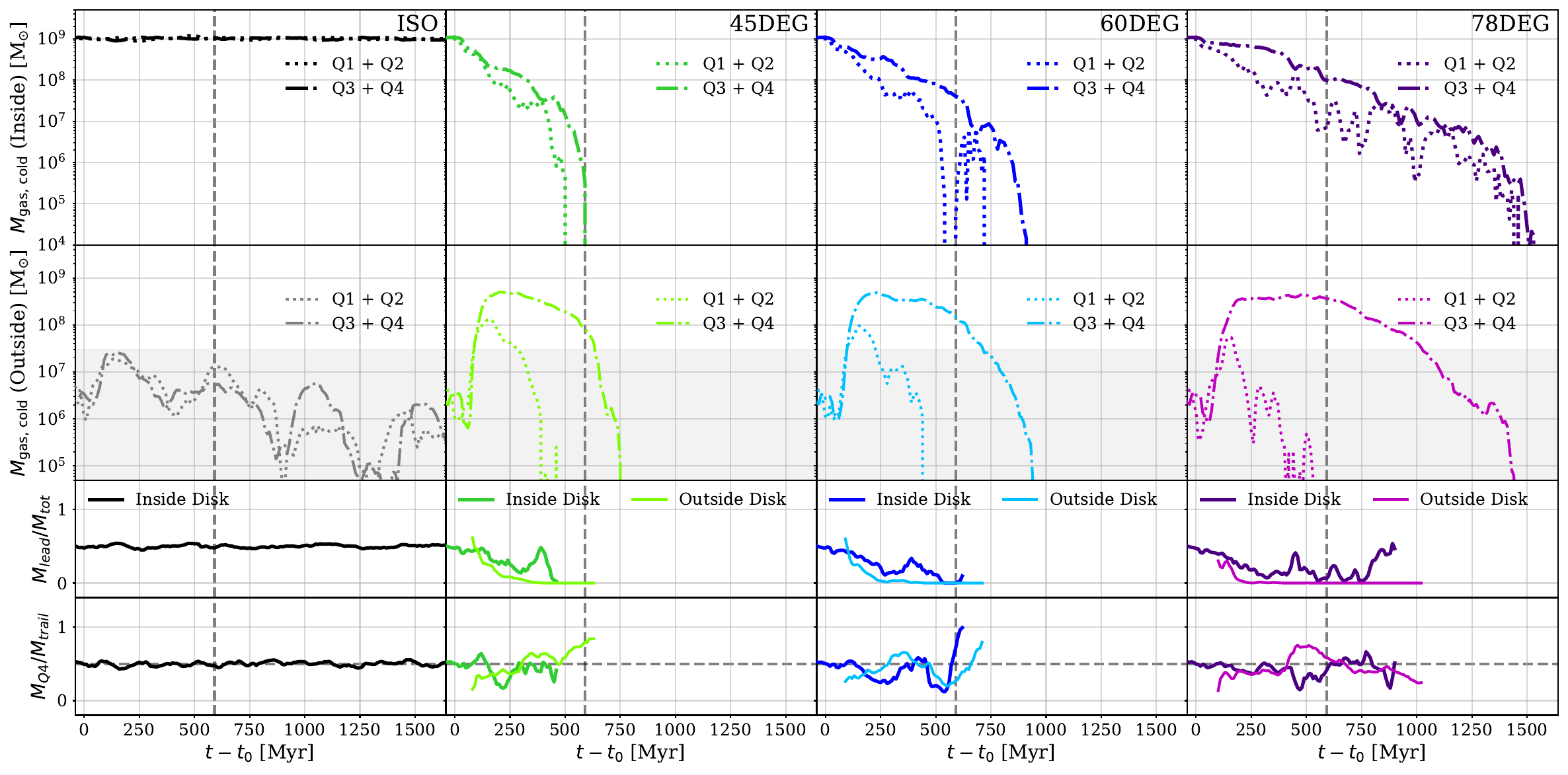}
    \caption{Quadrant mass evolution for the 4 simulations. The profiles shown are for cold gas \tcold{} only. Rows 1 and 2 show mass versus time for the leading (Q1 and Q2) and trailing (Q3 + Q4) sides, for inside and outside of the static disk, respectively. The 3rd row shows the ratio of masses on the leading side  compared to the total for inside (darker lines) and outside (lighter lines) the disk. Similarly, the 4th row shows the fraction of cold gas on the trailing side of the galaxy that is in Q4. 
    Vertical dashed lines indicate the point of peak ram pressure. 
    To only show ratios  when the total mass budget is sufficiently high, we only plot values in the third and fourth rows if the total available gas mass (in that region) is greater than the threshold level shown as the shaded gray region in the second row. This value is set to match the peak cold gas mass outside of the disk in the ISO run.
    }
    \label{fig:quadrant-mass-evol_cold}
\end{figure*}

With an inclined wind, global measurements do not tell the full story of how ram pressure affects a gas disk. 
Therefore, we examine the gas in the four quadrants introduced in Section \ref{sec:quadrants-criteria}, and how the quadrants can reveal asymmetries in the gas distribution that are amplified with increasingly inclined winds.

Figure \ref{fig:quadrant-mass-evol_cold} shows the evolution of cold gas in each of our runs, with the gas subdivided into these quadrants. Each column shows the results of one of our simulation runs.
From top to bottom, we show evolution of the cold gas mass (divided into the combined leading-side (Q1 + Q2) and trailing-side (Q3 + Q4) quadrants) first inside the static disk (defined in Section \ref{sec:selection-criteria}), and then outside of the static disk. 
For clarity, and because the leading-side quadrants show little asymmetry, we do not plot the quadrants individually in the first two rows.
We then look at the fraction of cold gas on the leading-side of the disk, and finally look at the fraction of trailing-side cold gas that is in Q4 specifically.
The time of the first pericentric passage is indicated in all panels with the vertical dashed line.

We set our expectations using the ISO run in the left column. Every panel shows that the ISO galaxy is quite symmetrical with little difference between the quadrants. Only in the second row, which shows the mass outside of the static disk, do we see some variation between quadrants. 
This is due to the stochastic nature of star formation and subsequent feedback, which will asymmetrically expel gas beyond the disk. We can use this as a mass threshold, approximately $10^7$ \msun, below which differences in mass across the quadrants can be explained by internal processes.
For rows 3 and 4 which involve ratios between quadrant regions, we only show the ratios when there is more than this mass in total.

Focusing on the top row, the three wind runs show gas removal in all quadrants, where the most inclined wind takes the longest to strip the ISM (also see Figure \ref{fig:disk-gas-mass-evol}).
In all wind runs, the leading-side quadrants (Q1 and Q2) always have less gas than the trailing-side quadrants (Q3 and Q4).
Variations between these pairs of quadrants are not as strong.

The second row shows the gas mass of each quadrant pair, but now outside of the static disk. Generally, we see the same behavior in the leading side quadrants for all wind runs. 
Within 400 Myr of the wind reaching the galaxy disk, the tail only extends in the trailing side quadrants in line with our expectations. 
After a steep initial increase, the trailing side mass outside of the disk plateau at $\sim5 \times 10^8$ M$_\odot$ for each quadrant pair. 
The trailing side quadrants then lose their mass, with the more inclined winds taking the longest time to lose their gas. The trailing side quadrants outside of the disk reach a state with only sparse cold gas ($< 10^6$\msun) either at the same time or slightly after all cold gas is removed from within the disk. 

Looking at the fraction of gas on the leading side in the third row, there is a steady decrease within the disk for all wind simulations up until the pericentric passage. In the 78DEG run, there are small, episodic increases in the fraction of cold gas on the leading-side within the disk (most prominently at $t-t_0=450$ Myr) that bring the ratio towards an even distribution on the leading and trailing sides. This effect appears to be the result of larger cold gas complexes orbiting around the center of the disk. 
By the time of the pericentric passage, much of the ordered rotational motion in the galaxy before the wind arrives has already been washed out.
Following the pericentric passage, there are fluctuations in the 78DEG run of the ratio between the leading and trailing sides within the disk between $0-60\%$, indicating some sloshing of remaining gas around the galaxy center. 
Outside of the disk, the ratio between the leading and trailing sides rapidly falls, although in all cases there is a small bump at around $t-t_0=75$ Myr which is probably due to stellar feedback around the time the wind is injected.  Once there is no mass on the leading side of the galaxy, it remains that way for the remainder of the simulation.

The fourth and final row shows the fraction of gas on the trailing side that belongs to Q4. 
For all simulations before pericenter, there appears to be a gradual trend (outside the disk) that begins with a low Q4 fraction at early times, growing to a larger Q4 fraction at later times.
However, this fraction appears to have significant variation over time, likely due to larger components of the ISM being stripped all at once into Q3.
The general trend is consistent with the evolutionary picture proposed in \citep{souchereauALMAJELLYHighResolution2025}, which argues that for an inclined wind, gas will be preferentially removed from the disk in Q3, but angular momentum will carry the gas into Q4 where it will accumulate and/or fall back into the disk.
Towards the end of each stripping event, fluctuations increase as the total amount of mass remaining in the galaxy (and therefore the mass available to be stripped) decreases. 
Both inside and outside of the disk, the trends are highly dynamic with time for all wind runs. This suggests that observations of the cold gas RPS galaxy tails may show very high variability, and any asymmetries might be short-lived.
We also note that our simulated galaxy has structure but no significant spiral arms, and the addition of spiral arm patterns into an RPS galaxy's morphology might act to increase these asymmetries.

Quadrant analysis is useful in that it is directly applicable to spatially resolved observations of RPS galaxies, so long as the on-sky position angle of the ram pressure wind can be measured. We have shown that (1) before pericenter, the leading side steadily loses mass compared to the trailing side, and (2) the two quadrants in the trailing side can develop high asymmetries that show strong variation with time.
Interestingly, the quadrant comparisons are quite similar for our three wind angles.
The main difference is that the stripping process is longer lived for nearly edge-on interactions, which would make observations of these asymmetries more likely in satellites experiencing more inclined winds.

\section{ Measuring Fallback}
\label{sec:measuring-fallback}

In this section we describe and then implement our methods to derive the spatial distribution and quantity of ISM fallback over time.
The first method, using projected distance-velocity phase space, is useful to spatially identify fallback.
The second method involves mass flow rates, and helps to quantify the amount of fallback over time.

\subsection{Identifying Fallback}
\label{sec:identifying-fallback}

\subsubsection{Identification in Distance-Velocity Phase Space}
\label{sec:ident-vtail-dtail}

\begin{figure*}
    \centering
    \includegraphics[width=\textwidth]{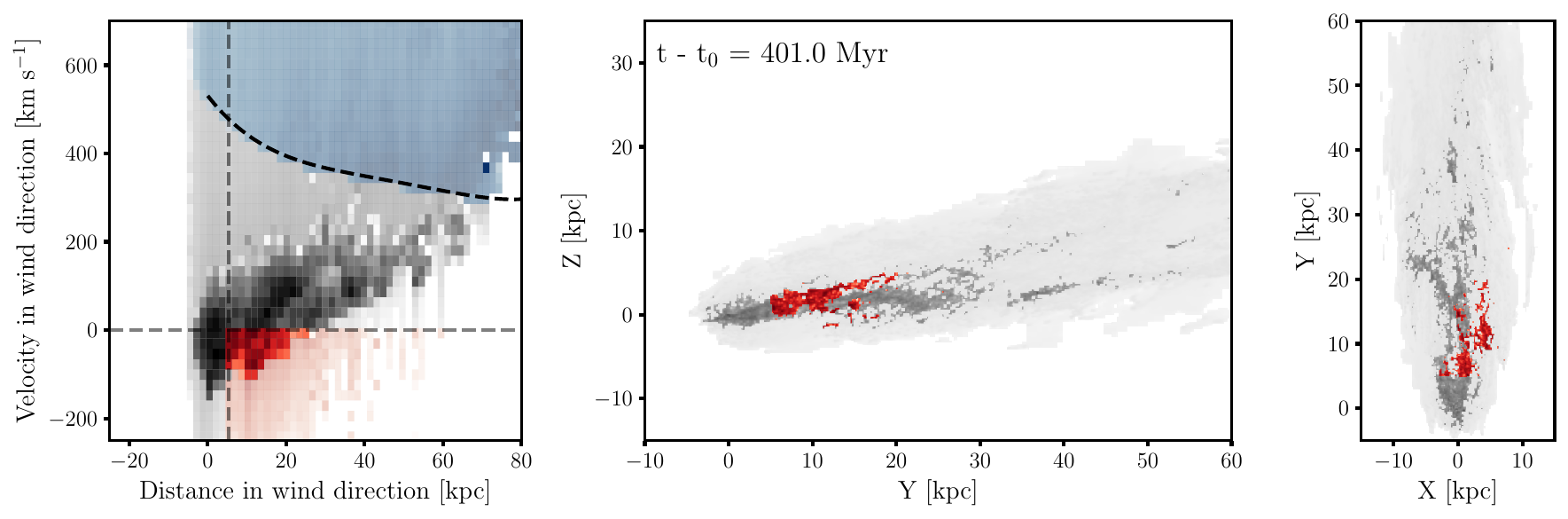}
    \caption{\textbf{Left:} \dtail-\vtail{} (Equations \ref{eq:dtail} and \ref{eq:vtail} ) phase plots for an example snapshot of the 78DEG run with high fallback (t = 431 Myr after the wind reaches the disk). The thick black dashed curve indicates the local escape speed as estimated from the gravitational potential. The vertical grey dashed line indicates the truncation radius (i.e. the current galaxy disk size) measured from this snapshot. Gas defined as fully stripped is colored blue, whereas gas defined as falling back is colored red. \textbf{Center and Right:} Gas density projections (side-view and top-down) of the same snapshot. ISM gas (gas with $Z/Z_\odot \geq 0.25$ with no temperature cut) is shown in light grey. Cold gas is shown in dark grey, and the cold gas defined as being in a fallback phase using Equation \ref{eq:fallback} is overlaid in red.}
    \label{fig:stfb}
\end{figure*}

\begin{figure}
    \centering
    \includegraphics[width=0.49\textwidth]{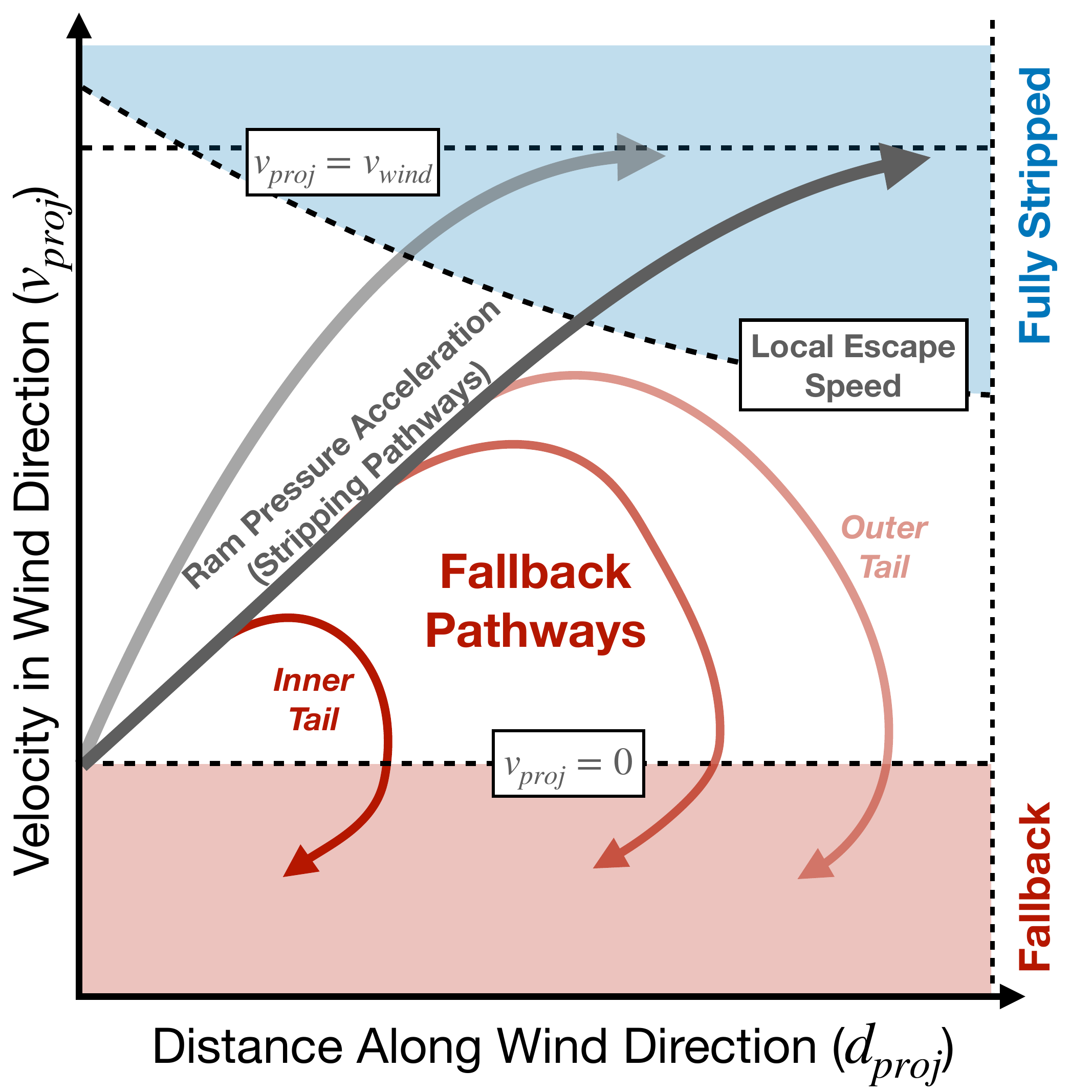}
    \caption{A schematic showing how gas evolves in the \dtail-\vtail{} phase space. Ram pressure accelerates gas along pathways towards the local escape speed. Darker arrows represent pathways that are more likely to be traversed by denser gas. Ram pressure loses efficiency as gas approaches the wind velocity (denoted $v_0$ in this diagram), but at greater distances from the galaxy center, the gravitational restoring force weakens as well. If the ram pressure acceleration is reduced or halted, gas might enter a ``fallback track" which carries gas back towards negative \vtail{} values.  Some examples of ``fallback tracks" are shown as red arrows. The gravitational restoring force becomes weaker at large \dtail, so most fallback would be expected close to the disk.}
    \label{fig:stfb-schematic}
\end{figure}

We seek to explore inner-gas-tail evolution such as fallback by examining the 6D phase space offered by our simulations, and to do so in a way that takes into account the geometry of the RP interaction. 
In particular, we look to find a phase-space that explicitly accounts for the ``disk-wind angle" $\phi_{DW}$, or the angle between the vector normal to the disk plane and the wind vector. $\phi_{DW}$ is $0\degree$ for face-on winds, and $90\degree$ for edge-on winds. 
To achieve this, we generalize the phase space representation  in \cite{tonnesenStarFormationRam2012}, which creates (for a face-on interaction) the distribution of gas and star particles in wind-direction velocity versus height above the disk. 
In their case, the wind vector is entirely in the $\hat{z}$ direction, so it is a $v_z - z$ phase space. We project the velocities and positions of gas clouds onto the wind vector, taking the center of the galaxy as the origin. This gives us two components, $d_{\text{proj}}$ and $v_{\text{proj}}$, which account for the disk wind angle and can be used to construct a physically-motivated definition for fallback.

In our simulations, our wind vector $\vec{v}_{\text{wind}}$ does not have an $\hat{x}$ component, and can be expressed in terms of the total velocity and the disk-wind angle as

\begin{equation}
    \vec{v}_{\text{wind}} = \begin{pmatrix} 0 \\ v_{\text{tot}}\sin(\phi_{\rm DW}) \\ v_{\text{tot}}\cos(\phi_{\rm DW}) \\  \end{pmatrix}.
\end{equation}

\noindent We generate two fields, the first being the projected distance along the tail, given by

\begin{equation}
    d_{\text{proj}}(y, z) = (y - y_0)\sin(\phi_{DW}) + (z - z_0)\cos(\phi_{DW})
    \label{eq:dtail}
\end{equation}

\noindent where the central $x$ and $y$ coordinates $x_0$ and $y_0$ are both 81 kpc (the centre of the galaxy in the simulation volume). The second field we derive is the velocity component in the wind direction, given by

\begin{equation}
    v_{\text{proj}}(v_y, v_z) = v_y\sin(\phi_{DW}) + v_z\cos(\phi_{DW}).
    \label{eq:vtail}    
\end{equation}

\noindent Notice that for a face-on interaction ($\phi_\text{DW} = 0$), this reduces to

\begin{align}
    v_{\text{proj}, 0} = v_z \hspace{2em} d_{\text{proj}, 0} = z - z0 = h
\end{align}

\noindent or the $v_z-h$ phase space as used in \cite{tonnesenStarFormationRam2012}. We use these two fields to derive criteria for gas at different phases of evolution. Gas that has been \textit{fully stripped} is given by

\begin{equation}
    |v_{\text{proj}}| > v_{\text{esc}}\left(d_{\text{proj}}\right)
    \label{eq:stripped-gas}
\end{equation}

\noindent where $v_{\text{esc}} = \sqrt{2|\Phi(d_\text{proj})|}$ is the local escape velocity determined from the gravitational potential. This is calculated along the mid-plane of the galaxy $(z = 0)$. 
Our definition of fallback requires two criteria:

\begin{equation}
    \begin{cases}
        v_{\text{proj}} < 0 \\
        d_{\text{proj}} > R_\text{trunc}(t) &
    \end{cases}
    \label{eq:fallback}
\end{equation}

\noindent where the first requires the gas be moving against the wind, and the second requires the gas be located outside of the region of the gas disk. While we believe that fallback likely persists for gas following reentry into the disk, it is not trivial to distinguish between disk gas moving against the wind due to fallback, or simply due to rotation. A more detailed study would require the ability to track gas clumps through simulation snapshots, which cannot be readily achieved with our \enzo{} setup.

Figure \ref{fig:stfb} shows the phase space distribution of cold gas \tcold{} and corresponding gas density projections in one snapshot of the 78DEG simulation in which fallback is at a high point ($t-t_0 = 431$ Myr). As expected from acceleration due to ram pressure, gas is pushed to both higher velocities and higher distances. 
Importantly, any of the gas in the grayscale region is still bound to the disk, so reducing the ram pressure from the wind could result in fallback. 
Indeed, we see this cycling of gas in phase space, where gas that is no longer being strongly accelerated eventually curls back in phase space towards the fallback region.
We include a time series of the evolution of gas in this phase space for the 78DEG run in Appendix \ref{app:projection-timeseries}.

We summarize this behavior in Figure \ref{fig:stfb-schematic}, which shows a schematic illustration of the \dtail-\vtail{} phase space. The total acceleration on the cloud is given by

\begin{equation}
    \vec{a}_{\rm cloud} =  \vec{a}_{\text{}ram} + \vec{a}_{\rm grav} = \frac{\rho_{\text{ICM}}|\vec{v}_{\text{rel}}|^2}{\Sigma_{\text{gas,cloud}}}\hat{v}_{\rm rel} - \nabla\Phi_{\rm gal}
    \label{eq:gunn-gott}
\end{equation}

\noindent where the ram pressure accelerates the gas towards both higher \vtail{} and \dtail{} values, becoming weaker as the relative velocity $\vec{v}_{\text{rel}} = \vec{v}_{\text{wind}} - \vec{v}_{\text{cloud}}$ approaches 0 as $\vec{v}_{\text{cloud}} \rightarrow \vec{v}_{\text{wind}}$. The gravitational acceleration also weakens at higher distances from the galaxy center. However, if at a point along this pathway ram pressure becomes weaker than the gravitational restoring force (for reasons we will discuss later in Section \ref{sec:why-fallback-occurs}), the gas could enter a fallback trajectory, where its velocity begins to decline, eventually reaching negative values of \vtail, and ultimately falling back towards the disk. 

This prescription may be challenging to apply to RPS galaxies with limited observational data, as kinematic data is generally restricted to line-of-sight motions, and positional information is mostly restricted to the viewing plane.
Despite this prescription requiring knowledge of the wind-disk angle in detail and the total velocity of the galaxy, it is a physically-motivated prescription that allows for direct comparison of galaxies with different wind inclinations.

\subsubsection{Identification Using Mass Flow Rates}
\label{sec:ident-mass-flow-rates}

While the first method to measure fallback provides a robust metric to spatially identify where in the tail fallback is occurring, we also seek a prescription that can help quantify the inflow rates which can then be compared to other physical processes.
For that purpose, we can turn to mass flow rates which take into account both the spatial and velocity information of the gas.

The mass flow rate of gas in and out of the main body of a galaxy can be measured by examining the velocity of gas in a spherical shell surrounding the ISM disk. This parameter has the benefit of being another physically relevant tracer of stripping and fallback, as it directly traces inflows and outflows. We can also change the size of the shell to explore the distances from the disk where fallback may be most dominant. Since we have simulations with multiple disk-wind angles, we choose to calculate the mass flow rate through a spherical shell of width 1 kpc. We do this for each snapshot by summing over the cells within the shell using

\begin{equation}
    \dot{M} = \frac{1}{R_{\text{eff}}}\sum_{\text{shell}}m_{\text{cell}}v_{\text{cell}}
    \label{eq:mass-flow-rates}
\end{equation}

\noindent where the effective radius $R_{\text{eff}}$ is computed by summing the volume of cells within the shell, and dividing by the surface area of the shell ($A_{\text{shell}} = 4\pi R^2$).



\subsection{Quantifying Fallback over Time}
\label{sec:quantifying-fallback}

\begin{figure}
    \centering
    \includegraphics[width=0.49\textwidth]{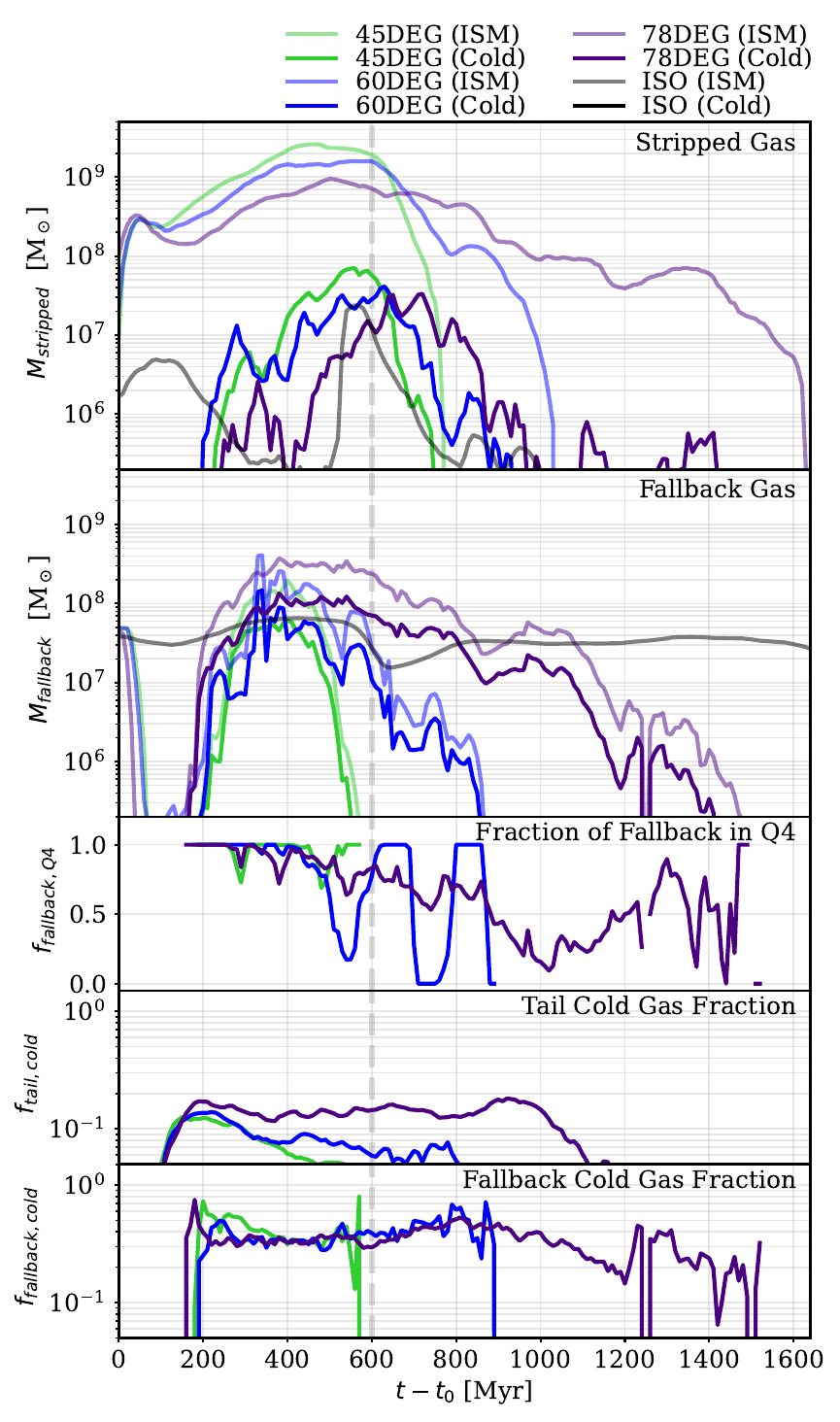}
    \caption{Time evolution of stripped and fallback mass using the \dtail-\vtail{} prescription described in Section \ref{sec:ident-vtail-dtail}. The first two rows show the fully stripped (blue region of Figure \ref{fig:stfb-schematic}) and fallback (red region of Figure \ref{fig:stfb-schematic}) gas masses over time. Fainter lines show ISM gas, whereas darker lines show cold gas.
    The third row shows the fraction of fallback gas located on the trailing side of the galaxy rotating into the wind (quadrant 4).
    The fourth row  shows the fraction of tail gas that is cold, while the fifth row shows the fraction of fallback gas that is cold.}
    \label{fig:stripped-fallback-evol}
\end{figure}

In the previous subsection we have defined two distinct methods of identifying fallback. We now apply these methods to our simulations to identify the amount of fallback at any given time, and explore gas inflow rates.

\subsubsection{Fallback Mass Over Time}
\label{sec:quantifying-fallback-dtail-vtail}

By summing the mass of gas based on where it resides in \dtail-\vtail{} phase space, we can quantify the amount of stripped and fallback ISM for each simulation snapshot. We show our findings in Figure \ref{fig:stripped-fallback-evol}. We include the ISO run (all ISM, in gray) to set expectations for what can be caused just by internal processes in the satellite galaxy. 
The ISO run has no fully-stripped cold gas, and only trace amounts of cold gas fallback.
The first and second panels show the amount of fully stripped and fallback gas over time (blue shaded and red shaded regions in Figure \ref{fig:stfb}, respectively). Light lines show all ISM gas, and dark lines show the cold ISM.

In the top row we see that at every time, cold gas is not a significant proportion of the fully stripped gas. This is expected as cold gas clouds have the highest surface densities and thus are the hardest to strip.
The decline in total ISM stripped gas is due to gas exiting the simulation volume, whereas for the cold gas, it can also be heated by the ICM wind.

The second panel shows that except for the earliest and latest times (approximately the first 200 Myr, and the final 50-100 Myr before becoming fully stripped), fallback is detected for the entirety of the RPS event.
In the ISO case, only warm ISM shows up as fallback. Even if stellar feedback can push some cold gas out of the disk (as seen in Figure \ref{fig:quadrant-mass-evol_cold}), only a small amount will be infalling at any given time.
Thus, it is clear that the ram pressure wind is required to strip cold gas to any appreciable level, as only diffuse ISM can be ejected (due to star formation feedback) in the ISO run. 
There is more fallback cold gas than stripped cold gas at any given time, which implies that most of the cold gas that leaves the disk ultimately falls back, or is ultimately heated through mixing with the ICM \citep[e.g.][]{tonnesenItsCloudsIllusions2021, choiRamPressureStripping2022}.

The third row shows the amount of cold fallback gas located on the trailing side quadrant rotating into the wind (Quadrant 4). It can be seen for all runs that at early times the fallback is almost entirely in Q4.
This indicates that rotation is a dominant driver of fallback at early times in our simulations. 
At later times in the 78DEG run post pericenter, the fraction of fallback in Q4 becomes more balanced with Q3, as the ISM disk no longer has proiminent rotation.

The fourth row shows the fraction of cold gas outside of the disk (defined as the static 15 kpc disk). The ISO case has a much lower cold gas fraction than the wind runs, which shows that RPS is required to remove cold gas from the disk. 
On average, the cold gas fraction of the ISO case within the disk is between 45-55\% (see Figure \ref{fig:disk-gas-mass-evol}, bottom panel).
There is a higher cold gas fraction in the 78DEG tail compared to the other runs, which might imply that heating is less efficient when the galaxy is highly inclined (i.e. the wind is interacting with a smaller amount of the galaxy's ISM at any given time). 

The fifth row shows the cold fraction of fallback gas over time. For the first 900 Myr, in all runs we see that the fraction of cold fallback gas is generally stable between $40-50\%$. Even though stripping continues beyond this time in the 78DEG run, we see a gradual decline in the fraction of cold fallback gas.
This is probably due to a combination of gas heating in the tail by the ICM wind, and a reduction in wind strength post-pericenter meaning that lower-density gas can more easily fall back.
The 78DEG run, however, reaches a longer-lived stripping state after pericentric passage, and the nature of the stripping being experienced by the 78DEG run at later times appears to be driven by turbulent viscous stripping \citep{nulsenTransportProcessesStripping1982, roedigerRamPressureStripping2008}. 
We posit that this is not seen in the 45DEG and 60DEG runs because the gas is rapidly stripped at or shortly after peak ram pressure. 

Comparing the final row with the fraction of cold gas, it is immediately apparent that cold gas forms a disproportionately large fraction of fallback gas given that it composes a very small percentage of the overall tail. We generally see cold gas fractions of $5-20\%$ in the tail, meaning that the cold gas fraction of fallback gas can be $\sim 2-10$ times higher. 
It is beyond the scope of this work to determine whether all the cold gas that is falling back was removed from the disk as cold clouds or cooled within the tail, although this is an interesting avenue for future work.

In summary, cold gas, which is generally denser and harder to efficiently accelerate, is more likely to fall back after being pushed out of the galaxy disk.  Because gas retains at least some of it's orbital motion, the fallback tends to happen in Q4.  We will discuss this in more detail in Section 5.

\subsubsection{Mass Flow Rates}
\label{sec:mass-flow-rates}

\begin{figure*}
    \centering
    \includegraphics[width=\textwidth]{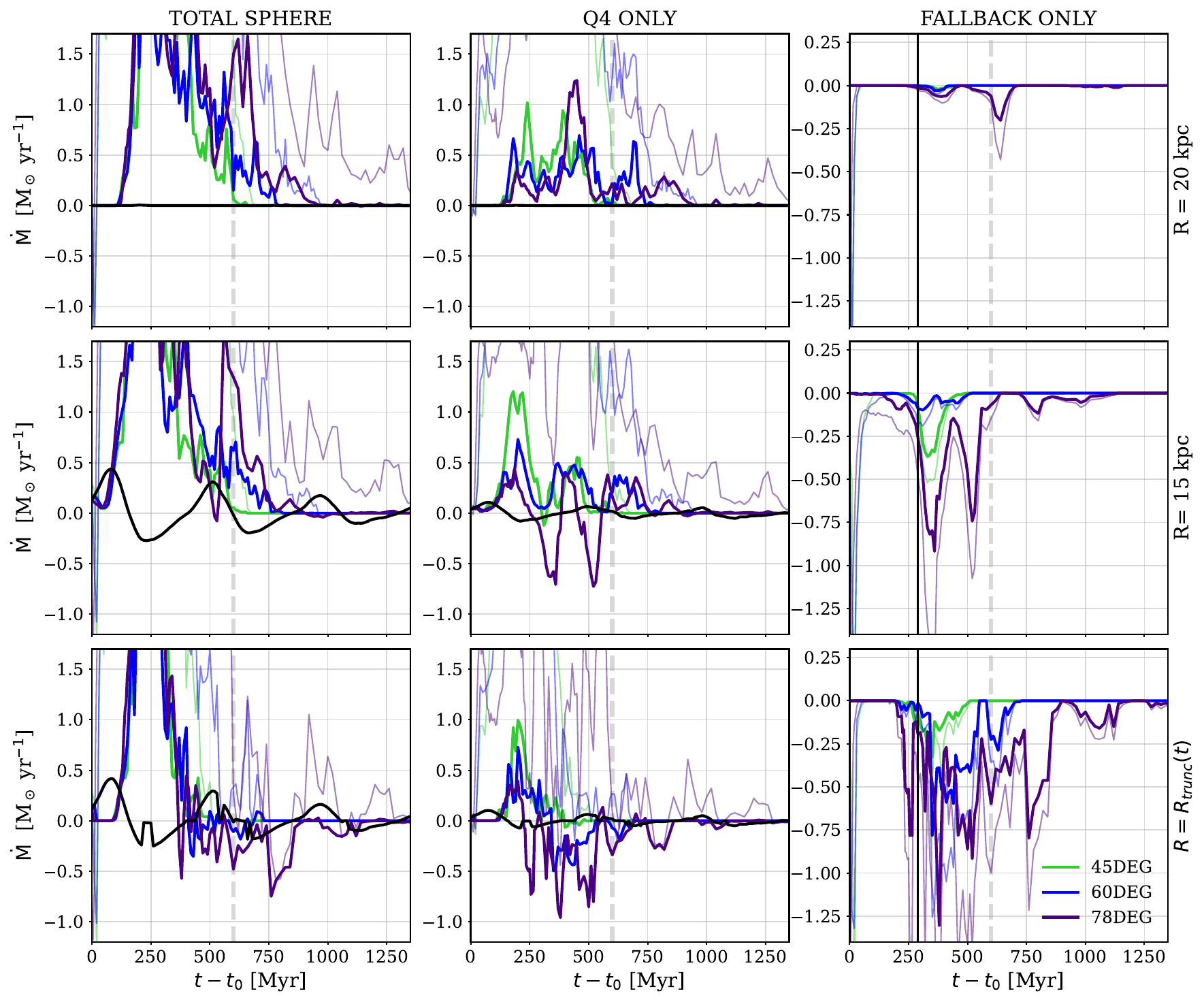}
    \caption{Measurements of the net mass flow rate (in M$_\odot$ yr$^{-1}$) of ISM gas (thin, light lines) and cold gas (thick, dark lines) through a spherical shell with different radii. We measure the mass flux through the entire sphere (left column) through only quadrant 4 (center column) and only the gas with negative $v_{r}$ values, (i.e. gas falling back, right column). Each row corresponds to a different shell radius, where from bottom to top the radii are $R=20$ kpc, $R=15$ kpc, and $R=R_{\text{trunc}}(t)$. The width of the spherical shell is 1 kpc for all shell radii. The vertical gray dashed line is the time of peak ram pressure. The vertical black solid line in the rightmost column is half the rotation period at 15 kpc for our satellite's gravitational potential.}
    \label{fig:mass-fluxes}
\end{figure*}

The first method of identifying fallback is most useful for spatial identification, and for examining how much fallback exists for an RPS galaxy at any given time. However, using mass flow rates can be useful to quantify the rate of infalling gas over time.
We show the results of our mass flow measurements in Figure \ref{fig:mass-fluxes}, where we plot $\dot{M}(t)$ (Equation \ref{eq:mass-flow-rates}) for spherical shells of different radii. 
We show the flow rates for the ISM gas (light, thin lines) and the cold gas (dark, thick lines). Each row shows the flow rates through a different shell radius, where from top to bottom, we show $R = 20$ kpc, $R = 15$ kpc, and $R = R_{\rm trunc}(t)$. In the first column we show the net mass flow rates through the entire sphere. 
In the second column we show the net mass flow rates but only through the quarter-sphere corresponding with Q4 (see Section \ref{sec:quadrants-criteria}). Lastly, in the final column we look at just the magnitude of the inflow rates.
We indicate the first pericentric passage with the vertical dashed lines.
While there are many interesting features in these mass flow rate plots, for this work we are focusing on what we can learn about fallback from these figures.

In the first column we can see that outflows dominate at all radii. This is in agreement with the negative slopes in the second panel of Figure \ref{fig:disk-gas-mass-evol}, showing gas loss in every simulation.
From this we see that the fallback rate beyond 15kpc is never larger than the stripping rate. 
However, when we look at the shell with $R = R_{\rm trunc}$,  there is some net inflow in the cold gas pre-pericenter in both the 60DEG and 78DEG runs.
We will discuss the likely mechanisms at work here in Sections \ref{sec:spatially-offset-rotation} and \ref{sec:shadowing}.
Post-pericenter, there is inflow in both the cold and ISM gas in the 78DEG run, which we will discuss more in Section \ref{sec:post-peak}.

Based on observations of ISM fallback in real RPS galaxies, we expect Q4 (trailing side, rotating into the wind), to have the most fallback \citep[see][]{souchereauALMAJELLYHighResolution2025}, so we look at the flow rates through that quadrant only in the second column.
At 20 kpc, we see no net inflows, suggesting that at a distance sufficiently far from the galaxy center, flows everywhere are dominated by stripping.
However, at 15 kpc, we do see episodes of net fallback for our most inclined case, but interestingly only for the cold gas.
Finally, at $R = R_{\rm trunc}$, we see net fallback in both the 60DEG and 78DEG runs. 
Pre-pericenter, the net fallback across the entire shell is less than that in Q4, but post-pericenter there is more net fallback across the entire shell than in Q4 only. 
There is also net fallback in Q4 earlier than there is in the entire shell.

We have thus seen so far that net fallback is rarer and the mass fluxes tend to be low. However, we can isolate the fallback across the whole shell to see when we get any fallback at all, which we show in the the final column. 
For all runs, there is at least a marginal amount of inflow at all radii, but fallback is highest closer to the boundary of the disk.
In the $R=15$ kpc shell, we see that the two large peaks of cold ISM fallback in the 78DEG run are similar to the net inflow rates in the Q4 quarter sphere.
However, in the Q4 region only cold gas shows net inflow and while there is ISM inflow shown in the final column, the net flows shown in columns 1 and 2 make it clear that the vast majority of ISM gas is out-flowing.
Especially at later times, most fallback consists of cold gas.

The mass fluxes support our results using the \dtail-\vtail{} method, that much more gas is out-flowing than is falling back, and cold gas composes a large fraction of fallback gas.
Here we have also seen that the majority of gas falling back does so within 15 kpc (i.e. the original extent of the disk), and very little gas that reaches beyond 20 kpc falls back.

\section{When and Why Does Fallback Occur?}
\label{sec:why-fallback-occurs}

We have defined and applied two methods to measure fallback in our simulation suite.
We now examine the processes that can give rise to fallback in ram pressure stripped galaxies.
To summarize our findings thus far, we have shown that:

\begin{enumerate}
    \item More fallback occurs at higher disk-wind angles. 
    \item Fallback predominantly occurs close to the remaining gas disk (i.e. in the ``inner tail").
    \item Fallback tends to occur on the trailing side of the galaxy on the side rotating into the wind (i.e. Q4 using our quadrant scheme).
     \item While cold gas is a small fraction of all gas in the tail, it constitutes a much larger fraction (up to an order of magnitude) of fallback gas.
    \item Fallback occurs during most of an RPS event (i.e. it is not a sporadic process). However, more fallback appears to occur before the pericentric passage.
\end{enumerate}

\begin{figure}
    \centering
    \includegraphics[width=0.5\textwidth]{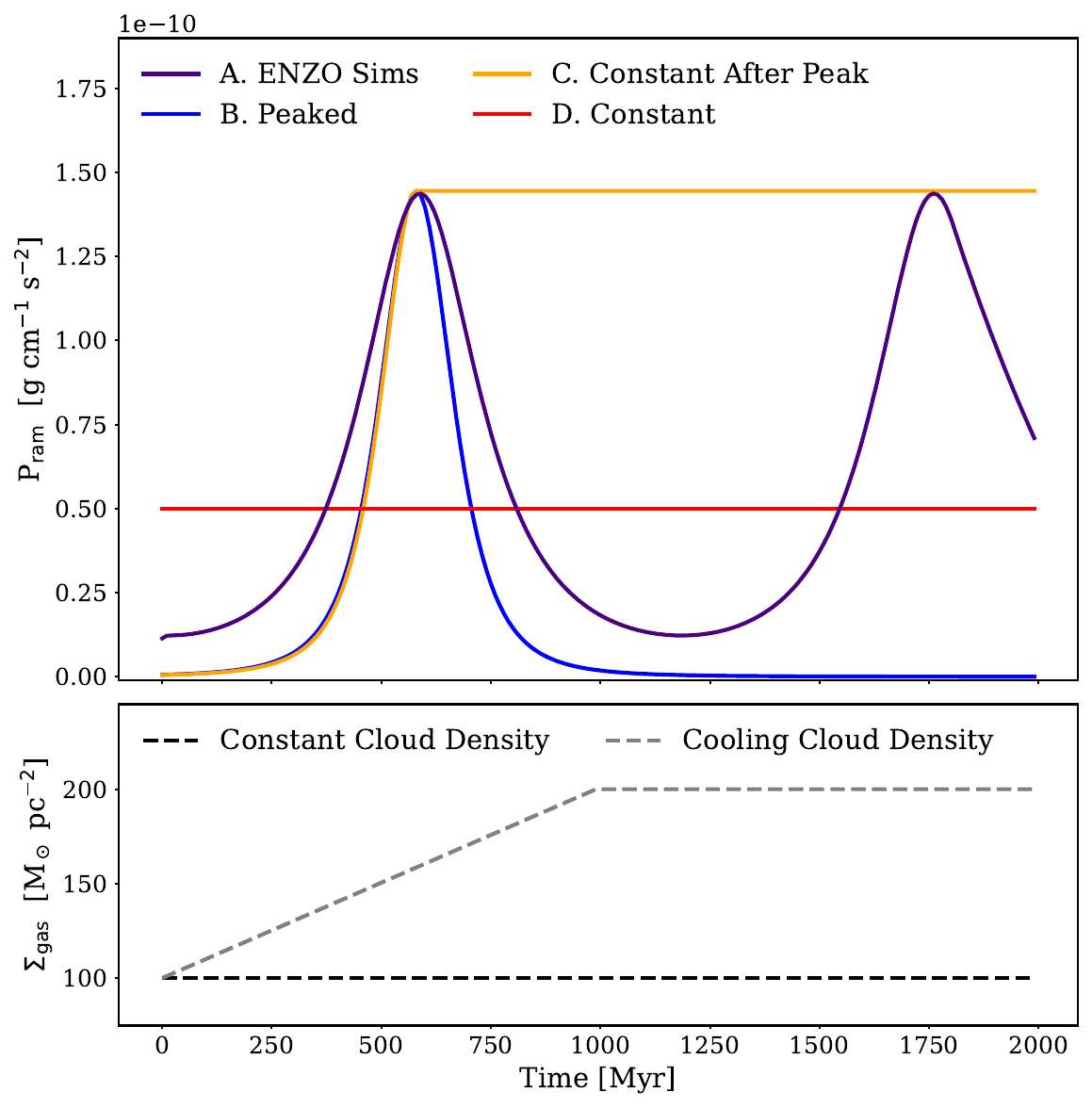}

    \caption{An overview of the idealized dynamical model setups. The top panel shows, as a function of time, the ram pressure wind strengths used, and the bottom panel shows the gas cloud surface densities.}
    \label{fig:wind-profiles}
\end{figure}

There are multiple possible explanations for why fallback can occur during an RPS event. In this section we discuss four such mechanisms that could cause fallback and illustrate how they can impact the trajectories of gas clouds.
The mechanisms are (1) spatially offset rotation caused by inclined ram pressure and rotation, (2) shadowing of the region downstream from the disk, (3) a decline in ram pressure strength following a pericentric passage, and (4) cloud cooling causing the surface density of a cloud to increase.

To demonstrate how each mechanism can contribute to fallback, we use a simple dynamical model built on the \textsc{Gala} package to model the orbits of a set of clouds orbiting through a smooth potential \citep{price-whelanAdrnGalaV1812024}. 
This model allows us to treat particles like individual gas clouds subject to ram pressure acceleration (in addition to the acceleration from the static gravitational potential) following Equation \ref{eq:gunn-gott}. The model shows the idealized effect of RPS on a gas cloud under a galaxy potential, without complications from (e.g.) gas self gravity, hydrodynamical instabilities leading to mixing, or radiative cooling. 
We also allow all of our gas particles in the disk to be ``pushed" by the wind, even if in a realistic simulation that inner gas would be protected by gas further out in the disk.

The strength of using this model is that we can track the motion of clouds over time to see where in a galaxy a dense cloud would be perturbed by interacting with a wind, and whether a perturbed cloud would be likely to escape or fallback. We configure the model to generate tuned examples that showcase each mechanism.
We don't intend for this dynamical modeling to be seen as a replacement or analogue for hydrodynamical simulations, as they lack much of the underlying physics required for detailed analysis of galaxy-wind interactions. However, the model can provide demonstrative examples for how each mechanism can manifest under idealized conditions of gravity, a ram pressure wind, galaxy rotation, and changes in cloud surface density.
 \footnote{As discussed in other works \citep[e.g.][]{tonnesenItsCloudsIllusions2021} as well as this paper, ram pressure can result in mixing that drives acceleration. In this dynamical model we assume that a cloud acceleration is given purely by drag $\left(P_{\rm ram} = \rho_{\rm ICM} |v_{\rm rel}|^2\hat{v}_{\rm rel}\right)$. This is one example of the simplicity of this model.}

\begin{figure}
    \centering
    \includegraphics[width=0.5\textwidth]{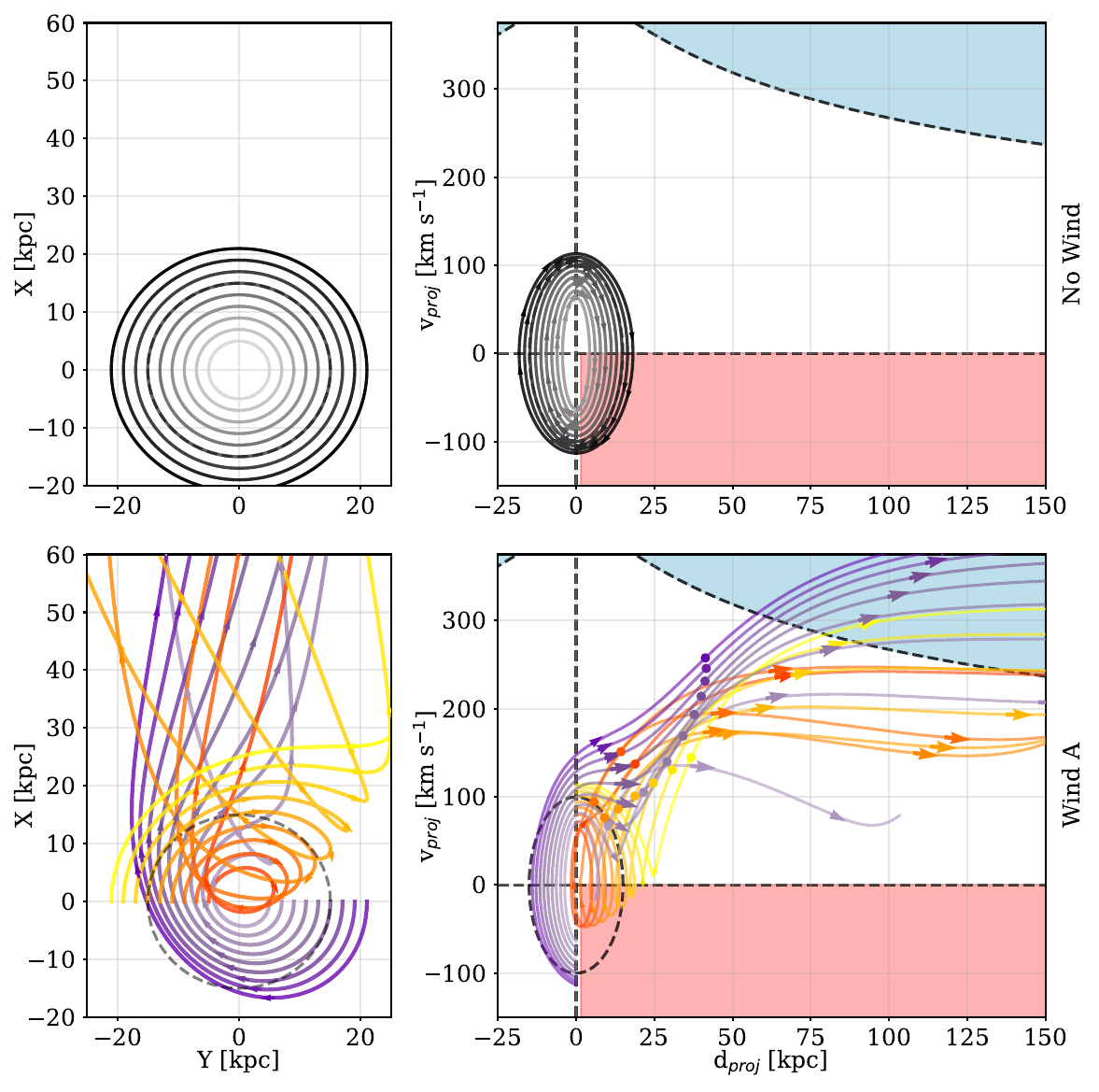}
    \caption{The orbits of test clouds through the same static potential as the \enzo{} simulations. The top row shows no wind applied (grayscale), whereas the bottom row shows the test clouds subjected to the same wind as our \enzo{} simulations (Wind A in Figure \ref{fig:wind-profiles}, colors). The left panels show an x-y (top down) projection of the clouds positions. The (projected) wind direction is vertical for these panels. The trajectories of the clouds are colored to coincide with their initial distance from the potential center. The red-yellow orbits are for clouds initially rotating with the wind, and the purple orbits are for clouds initially rotating against the wind. A dashed circle with $R=15$kpc is included to indicate the size of the unperturbed galaxy disk. The right column shows the \dtail-\vtail{} phase-space plots as discussed in Section \ref{sec:ident-vtail-dtail}. The region where clouds have become fully stripped is shown in blue, and the region where clouds are moving against the wind (fallback) is shown in red. The black dashed oval indicates the boundary of the region in phase space occupied by the clouds without any ram pressure (i.e. on unperturbed circular orbits). Arrows (spaced at 400 Myr intervals) indicate the direction of motion of the clouds, and the filled circles indicate the locations of clouds at the time of pericentric passage ($t \approx 600 $Myr).}
    \label{fig:inward-outward-orbits}
\end{figure}

The wind profiles and cloud surface densities explored in the following dynamical models are summarized in Figure \ref{fig:wind-profiles}. Unless otherwise specified, we use a wind that matches the one experienced by the galaxy in our \enzo{} suite (Wind A). We also examine a wind that follows a Lorentzian-like rise and fall similar to the \enzo{} suite but with only one peak (Wind B), a wind that rises to the same peak as Wind B but remains at the peak strength (Wind C) and a wind that remains at a constant intensity throughout (Wind D). For the constant wind, we adopt an ICM density of $\rho_{\rm ICM} = 5  \times 10^{-27}$ \gcmthree{} and a wind speed of 1000 \kms. 
All of the winds discussed in the following sections have a disk-wind angle of $\phi_{DW} = 78\degree$, consistent with the 78DEG run. 
For the cloud surface densities, we explore two different types of clouds: those with unchanging surface densities of 100 \msunpctwo, and those that gradually increase to 200 \msunpctwo{} over the duration of 1 Gyr, and then remains constant. 
Our choice of cloud surface density is motivated by the distribution of molecular cloud surface densities measured using $^{12}$CO observations \citep{miville-deschenesPhysicalPropertiesMolecular2017}. Since these models are purely idealistic and serve as demonstrations of the mechanisms that can generate fallback, we choose to leave the initial surface densities unchanged as a function of radius.
Each run adopts 100 \msunpctwo{} as the starting surface density for all clouds.

The gravitational potential is the same as the static potential in our \enzo{} simulation suite, which consists of the dark matter halo and stellar disk (See Table \ref{tab:potential}). 
Although clouds can be distributed arbitrarily through a disk, we are using these models to understand how clouds are disturbed by a wind, and whether they will fall back to the disk. Therefore, for clarity, the clouds are placed in two rows, with one row on the side rotating into the wind, and the second row rotating with the wind. They are placed at increasing galactocentric radii, spaced 2 kpc apart from $R=5$ kpc to $R=20$ kpc, and given initial velocities to have circular orbits. Like the \enzo{} simulations, the clouds are orbiting in a clockwise motion.

We use two projections to illustrate the motion of these clouds: an x-y (top-down) projection, and a \dtail-\vtail{} projection.
The top row of Figure \ref{fig:inward-outward-orbits} shows the clouds without any ram pressure applied. In the absence of a wind, the clouds remain on circular orbits. They also occupy a distinct elliptical region in the \dtail-\vtail{} phase space shown in the right column, which we indicate in all remaining rows with the dashed ellipse.

We discuss four plausible mechanisms in order of decreasing dependence on disk-wind angle. In the following subsections, we briefly describe each mechanism, and demonstrate using our dynamical model how each of these processes can contribute to ISM fallback.

\subsection{Spatially Offset Rotation}
\label{sec:spatially-offset-rotation}

As discussed in \cite{souchereauALMAJELLYHighResolution2025}, the development of an asymmetric molecular gas tail with an overabundance of gas in Q4 (trailing side, rotating into the wind) can be explained by the combined effects of galaxy rotation and torques from the ram pressure wind on the ISM.
Assuming the simple case of an edge-on wind with constant strength, gas on the side of the galaxy rotating \textit{with} the wind is preferentially stripped from the disk, as torque on this side of the galaxy increases the cloud's angular momentum, assisting in ram pressure stripping.
For gas rotating against the wind, torques from ram pressure drive gas radially inwards towards the potential center.
The gas clouds that have been pushed out of the galaxy disk, but have not yet reached the local escape speed have been accelerated onto an elongated orbit. 
The angular momentum of these clouds carry them across the width of the tail into Q4, where (assuming they remain on a bound orbit), the clouds will now be orbiting back towards the galaxy center.

In the second row of Figure \ref{fig:inward-outward-orbits}, we show the effects of the \enzo-like wind (Wind A in Figure \ref{fig:wind-profiles}) on gas clouds with unchanging surface densities of $100$ \msunpctwo.
In the top-down view in the left panel, clouds initially rotating with the wind (colored orange-red) are first pushed out of the galaxy disk and then move across the width of the tail into Q4.
The right panel shows the evolution of these clouds in \dtail-\vtail{} space (see Section \ref{sec:ident-vtail-dtail} and Figure \ref{fig:stfb-schematic}). 
For the clouds initially rotating with the wind, the ``cycling" effect is very clear with most of the clouds following a loop-like trajectory through the phase space. 
This includes some orbits that dip into the fallback region (shaded red) outside of the region of phase space occupied by the orbits when no ram pressure is applied.

For the clouds initially rotating into the wind, the top-down view shows that they are initially driven radially inwards, which can clearly be seen by comparing to the dashed circle with $R=15$ kpc. Following this inward motion, the combined effects of ram pressure acceleration and their increased angular momentum (due to the ram pressure torque) drive them out of the galaxy disk. 
By the time these clouds have orbited 180 degrees, (\dtail{} $ = 0$) they have been pushed to higher values of \vtail{} than the clouds that are initially rotating with the wind. 
This can be seen when comparing the values of \vtail{} for the clouds when they cross the dashed line at \dtail$=0$ in the right panel.
At this point their rotational motion has been overcome by the ICM wind's momentum, and the clouds follow ``stripping pathways" with no looping motions.

With these results, we find that when only considering the combined effects of a ram pressure wind, galaxy rotation, and angular momentum, it is possible to achieve \textit{some} degree of fallback.
This mechanism will be important only when rotational motion is still prominent in the galaxy.
We note that this effect can only occur when the ram pressure is not strong enough to rapidly remove all of the disk gas, and when there is a highly inclined wind-ISM interaction, since the wind must affect the angular momentum of the disk gas.

\subsection{Downstream Reductions in Local Ram Pressure Strength}
\label{sec:shadowing}

\begin{figure}
    \centering
    \includegraphics[width=0.5\textwidth]{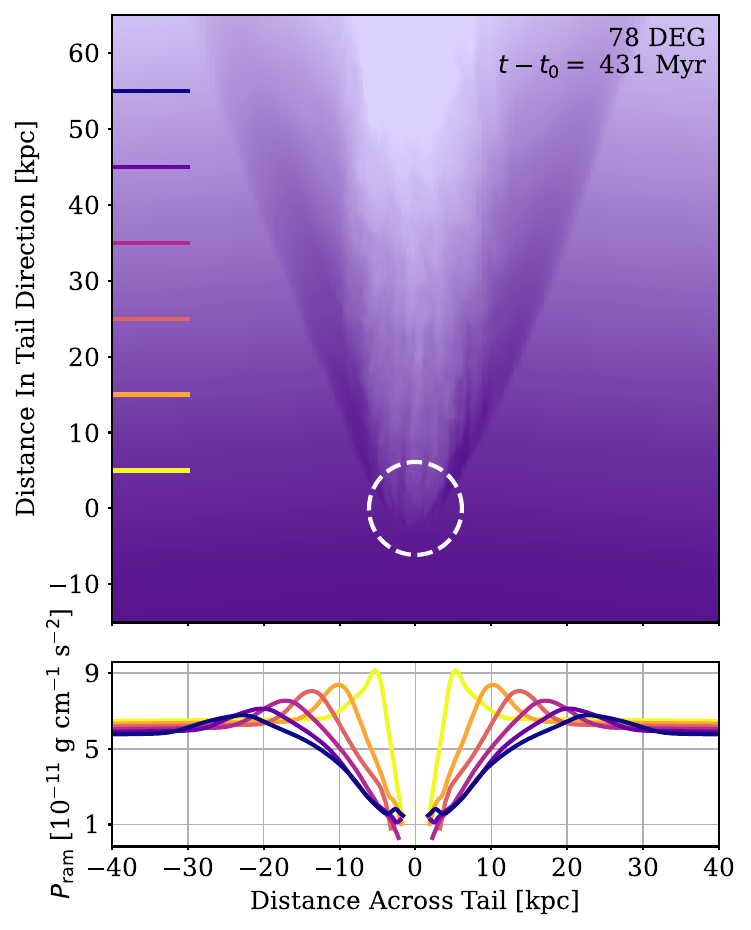}

    \caption{A measurement of the ram pressure shadow from one of the snapshots of the 78DEG simulation run (at $t-t_0 = 700$ Myr),  The top panel shows a projection of $P_\text{ram} = \rho v^2$ for gas with $Z/Z_\odot < 0.25$, where darker colours indicate stronger values of $P_\text{ram}$. The white dashed circle indicates $R_\text{trunc}$ for this snapshot. The bottom panel shows ram pressure strength profiles at different distances along the tail, indicated by the horizontal ticks in the top panel.}
    \label{fig:shadow-profiles-example}
\end{figure}

\begin{figure}
    \centering
    \includegraphics[width=0.5\textwidth]{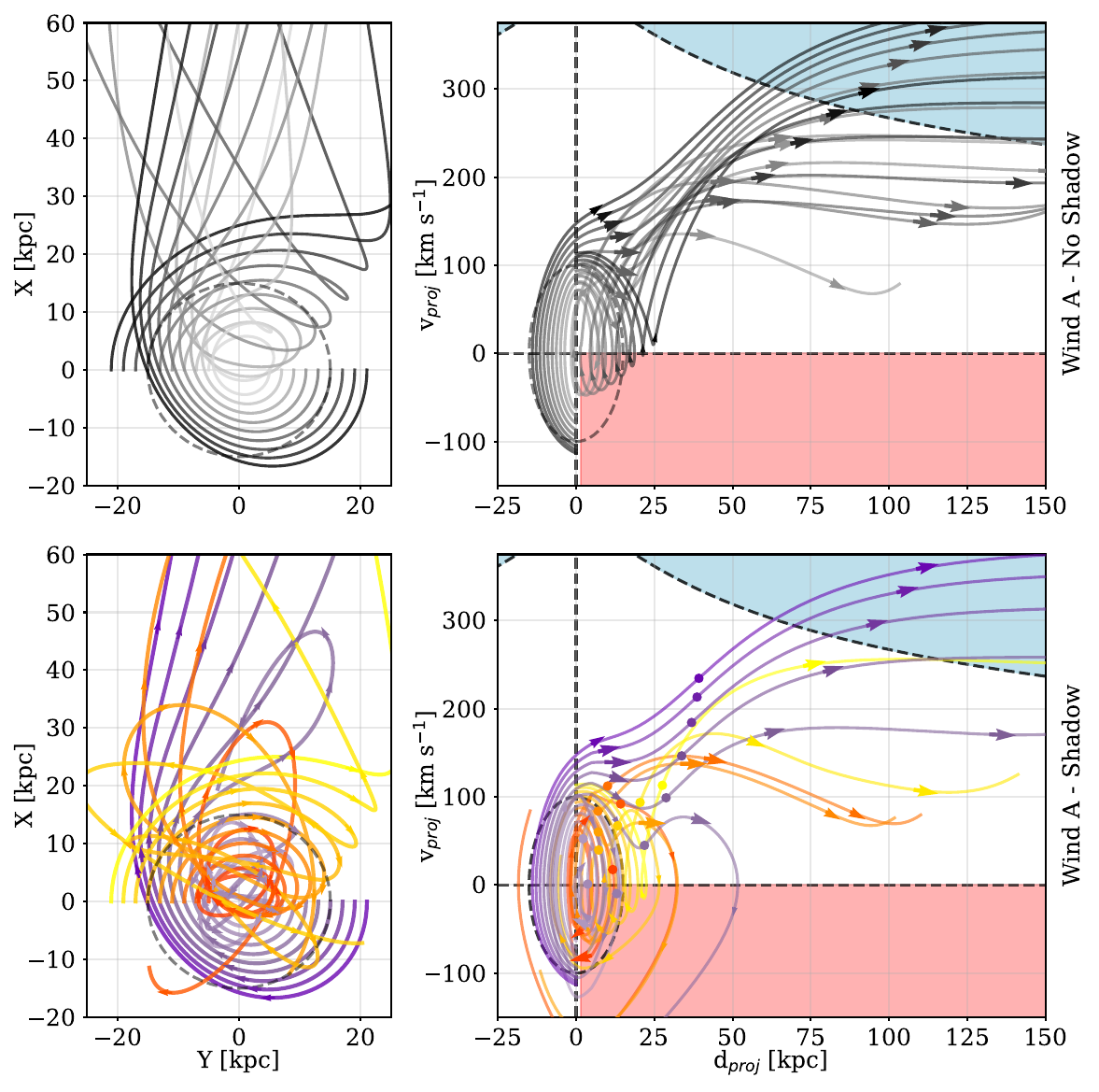}

    \caption{The same as Figure \ref{fig:inward-outward-orbits}, but with a shadow applied to reduce the efficiency of ram pressure acceleration downstream from the galaxy disk. The top row (grayscale) shows the orbits subjected to the wind without a shadowing, and the bottom row (colors) shows the orbits with the shadow applied.}
    \label{fig:inward-outward-orbits-shadowing}
\end{figure}

While we have shown that spatially-offset rotation can generate fallback on its own, it is likely that a reduction of ram pressure strength downstream from the remaining galaxy disk can also be a driver of fallback, including at times when disk rotation is not prominent.
This effect, called shadowing, has been identified before in other  simulations \citep[e.g.][]{vollmerRamPressureStripping2012, roedigerStrippedEllipticalGalaxies2015}. 
A ram pressure shadow can manifest through the wind running into the disk and being ``blocked" directly downwind of the galaxy.
It would naturally be expected, in the case of more inclined winds, that more stripped gas will pass into the shadow as it orbits within the tail. As an explicit example, in the case of a face-on wind, rotation does not act to carry clouds into the shadow as it does for inclined cases.

Although not the same mechanism, similar effects can be achieved through a bow shock if the galaxy is traveling supersonically.  
Bow shocks have been identified in observations of ram pressure stripping \citep{stevensGalaxiesClustersObservational1999b, bellhouseGASPIIMUSE2017}, as well as in hydrodynamical simulations \citep{yunJellyfishGalaxiesIllustrisTNG2019, gollerJellyfishGalaxiesIllustrisTNG2023}.
Furthermore, the shock cone created from a bow shock can be considerably wider than the remaining ISM gas disk, meaning that the entirety of a galaxy's RPS tail may be affected.
Because our galaxy is traveling supersonically through the massive host halo $(\mathcal{M} \approx 3-5)$, the bow shock is likely to be the principal mechanism creating downstream ram pressure reductions.

To determine the shape and extent of the shadow for our idealized dynamical models, we can return to our \enzo{} simulations and measure density and velocity for gas with $Z/Z_\odot < 0.25$. 
This selection removes most of the ISM gas but leaves enough cells to measure the local flow strength throughout the simulation volume. 
Figure \ref{fig:shadow-profiles-example} shows the structure of the ram pressure strength for a snapshot of the 78DEG run ($t-t_0 = 431$ Myr, the same snapshot as Figure \ref{fig:stfb}).
We see evidence for a bow shock that creates an over-dense region that widens well beyond the extent of the remaining gas disk. 
Near the symmetry line of the image is a drop in ram pressure strength that extends along the length of the tail to the end of the simulation volume. 

At a series of distances along the tail we also measure radial profiles of the ram pressure strength in slabs with widths of $\pm5$ kpc, tilted such that their normal vectors are pointed in the wind direction. The profiles, shown in the bottom panel of Figure \ref{fig:shadow-profiles-example}, all follow a similar trend, with the profiles flat at the outskirts, before becoming elevated at the ``edge" of the shock front, then dropping in strength towards the tail centerline. The elevated bumps are strongest closer to the galaxy disk, and the decline in ram pressure strength is steepest.
We confirmed that the elevated bumps are due to a gas over-density (rather than a velocity jump) being created by a bow shock, which factors into the measured ram pressure strength. As the shock dissipates downstream, the bumps become less severe.

Motivated by the ram pressure profiles we see in our \enzo{} simulations, we emulate a downstream wind reduction in our dynamical model by considering the distance each cloud is located from the center of the tail $r_\perp$. We can define the reduction of the wind strength, by way of the wind speed, using a Gaussian profile as

\begin{equation}
    v_\text{wind, reduced} = v_\text{wind} \left[1 - a \exp\left(- \frac{r_\perp^2}{2\sigma^2} \right) \right]
\end{equation}

\noindent where for our example we set $\sigma = 10$ kpc, and $a = 0.8$, meaning that in the center of the shadow, the wind strength is reduced by $80\%$. \footnote{Based on the steep decline of the ram pressure strength near the tail central axis in Figure \ref{fig:shadow-profiles-example}, this is likely a conservative estimate.} 
To allow for clouds to escape the disk region, we also require that the cloud be at least 5 kpc downstream (i.e. $d_\text{proj} > 5$ kpc) in order to be shadowed.
For simplicity, we do not include any increases in ram pressure at the edges of the bow shock. 
This modified wind speed is then applied to Equation \ref{eq:gunn-gott} to calculate the ram pressure acceleration on our clouds.

Figure \ref{fig:inward-outward-orbits-shadowing} shows the effect of reducing ram pressure with a shadow and/or bow shock on enhancing fallback while experiencing the same wind as the \enzo{} simulations. 
Unlike in the shadow-free regime where fallback is limited and all the clouds are ultimately stripped, the shadow allows for fallback throughout, and looping orbits can be seen which reach distances of 30-45 kpc downstream before returning to the disk. 
Clearly, including the reduction of ram pressure from a shadow and/or a bow shock increases the amount of gas that can fall back to a galaxy.

\subsection{A Global Decline in Ram Pressure Strength Post-Pericenter}
\label{sec:post-peak}

\begin{figure}
    \centering
    \includegraphics[width=0.5\textwidth]{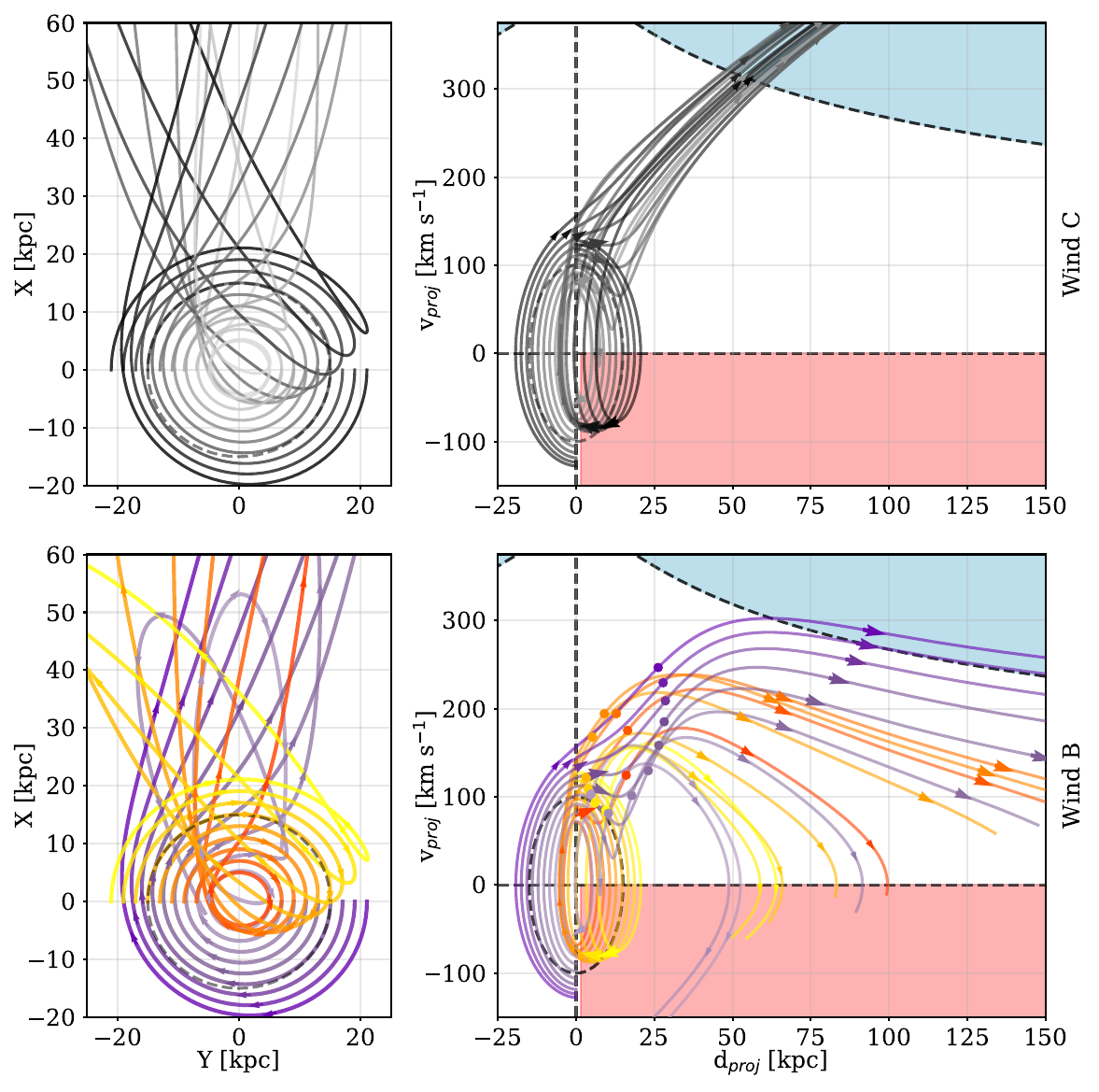}

    \caption{The same as Figure \ref{fig:inward-outward-orbits}, but comparing a peaked wind (Wind B - colored trajectories) to a wind that rises in an identical fashion, but then remains at peak strength (Wind C - grayscale trajectories).}
    \label{fig:orbits-post-peak}
\end{figure}

While the first two mechanisms cover local effects, the variable strength of a ram pressure wind on a global scale can also have an impact on fallback. The plunging nature of the orbits of galaxies through clusters creates peaked ram pressure strength profiles centered at the pericentric passage. 
This is true of our \enzo{} simulations, with a ram pressure profile modeled based on a possible orbit of NGC 4858 through the Coma cluster.
While the wind may grow to a point sufficient to strip gas from the galaxy, the decline post-peak might result in gravitational forces overcoming ram pressure, resulting in fallback.

To test the effects of a post-peak decline, we compare two idealized wind profiles. Both winds follow a Lorentzian-like increase, but their behavior differs after reaching peak wind strength. The first wind declines symmetrically (B in Figure \ref{fig:wind-profiles}), but the other (C in Figure \ref{fig:wind-profiles}) remains at peak ram pressure (similar to \cite{zhuWhenHowRam2023}). Figure \ref{fig:orbits-post-peak} shows the results of the dynamical model using these two wind profiles. For a wind that does not decline, we see complete stripping of all our modeled clouds (shown in grayscale). However, the wind reducing in strength is sufficient to see dramatically different effects, including many orbits beginning to re-enter the fallback region (at distances up to $d_\text{proj} \approx 100$kpc).

As long as the infalling galaxy has not been fully stripped by the time of the pericentric passage, the drop in ram pressure is likely sufficient to sustain fallback. This is most likely the dominant mechanism for the 78DEG run following peak ram pressure, which has fallback throughout this period despite only having $<10\%$ of its initial ISM reservoir remaining, and little to no rotational motion present.

Our simulation suite features a very strong ram pressure wind for a galaxy at this mass scale ($P_\text{ram,peak} \approx 1.5\times10^{-10}$ \rpunits{}). 
For example, this is approximately an order of magnitude higher than the strongest wind explored in \cite{zhuWhenHowRam2023} ($P_\text{ram,peak} \approx 1\times10^{-11}$ \rpunits{}). If a weaker ram pressure is able to accelerate gas from the disk but not drive it to escape velocity before pericentric passage, we predict that fallback could be elevated. 


\subsection{Cloud Surface Density Evolution}
\label{sec:cloud-compression}

\begin{figure}
    \centering
    \includegraphics[width=0.5\textwidth]{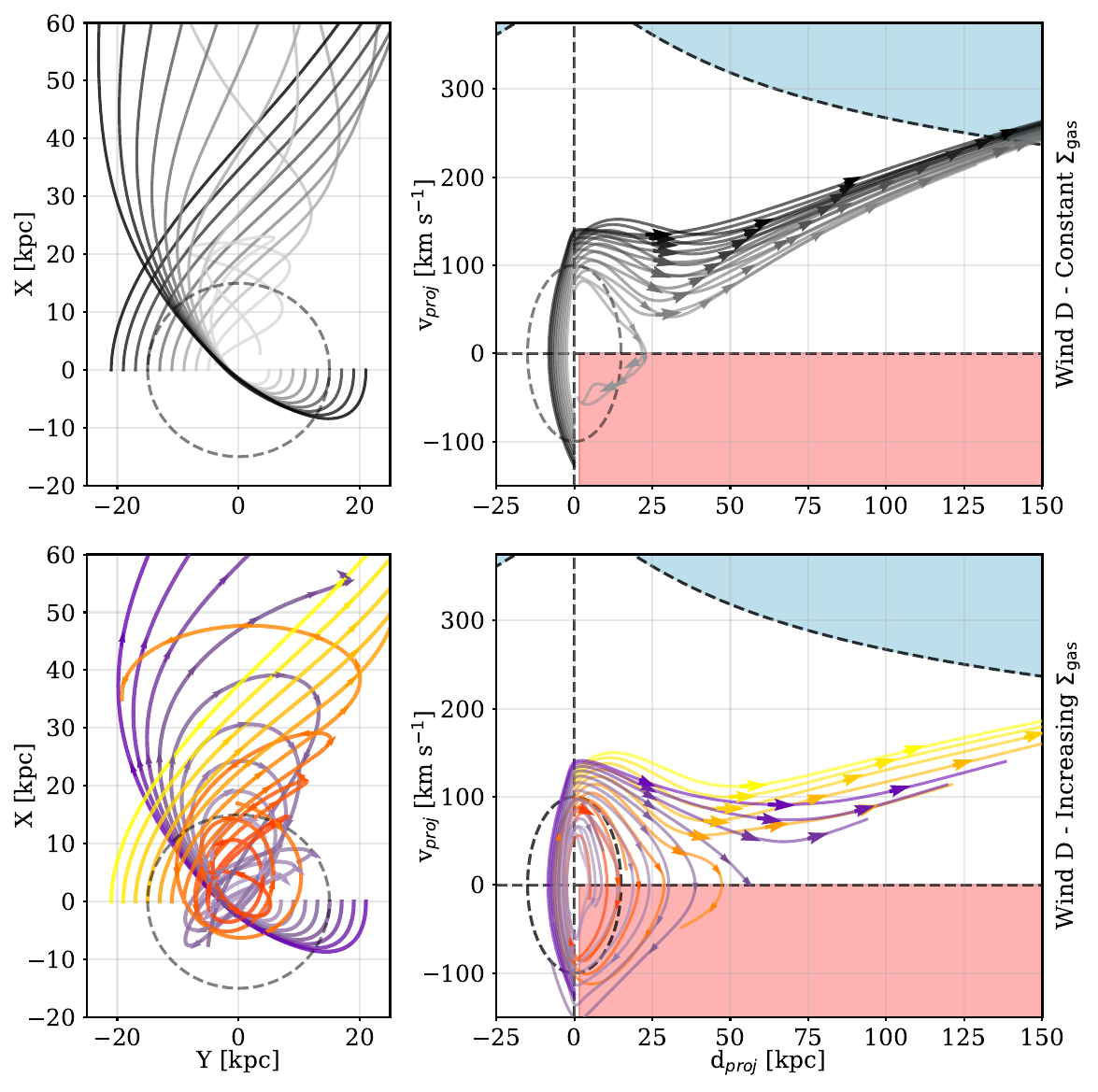}

    \caption{The same as Figure \ref{fig:inward-outward-orbits}, but comparing under the effects of a constant wind (Wind D in Figure \ref{fig:wind-profiles}) clouds that remain at 100 \msunpctwo{} (grayscale) to clouds that start at 100 \msunpctwo{} but steadily increase their surface density to 200 \msunpctwo{} over 1 Gyr (colors).}
    \label{fig:orbits-cloud-cooling}
\end{figure}

So far we have assumed that the gas clouds in our idealized dynamical models have remained at a constant surface density throughout. In this section, we consider the effects of an evolving gas surface density on the orbits of clouds under ram pressure. 
There are many ways that clouds might evolve to a larger surface density, including cooling, compression, and/or mass-accretion.
For this exercise, we will focus on a general increase in surface density without examining the underlying mechanisms in detail.
That being said, is already established that cold clouds can gain mass as they interact and mix with an external wind \citep[e.g.][]{gronkeHowColdGas2020, fieldingStructureMultiphaseGalactic2022}. 
If that mass growth outpaces the growth of the cloud's effective cross-sectional area, the surface density will increase and the cloud will become harder to accelerate.

To explore the effects of cloud evolution, we create two different dynamical models where the first set of clouds has their surface densities fixed at 100 \msunpctwo, and the second set of clouds starts at 100 \msunpctwo{} but steadily increases to twice that amount over the first 1000 Myr, and remains constant thereafter. 
We use the constant wind (Wind D in Figure \ref{fig:wind-profiles}) for both models.
If we assume that the size of the cloud remains approximately constant, we can see from Figure 4 of \cite{fieldingStructureMultiphaseGalactic2022} that it is reasonable to assume a cloud can double in mass over 1000 Myr.
Furthermore, cooling times for our clouds of interest can be orders of magnitude smaller than 1000 Myr \citep[See Figure 2 of][]{abruzzoSimpleModelMixing2022}.
Therefore, the surface density evolution we explore in the idealized dynamical modeling can be seen as a modest to conservative growth rate.
If the cloud grows in surface density faster than what we explore here, the effects on fallback will be even greater \footnote{For example, if the mass doubling time is 100 Myr as opposed to 1 Gyr, only the dense cloud orbiting with the wind at 20 kpc from the disk is pushed to beyond 50 kpc from the disk.}.

When examining the behavior of the two sets of orbits in Figure \ref{fig:orbits-cloud-cooling}, we can clearly see that gradual increases in surface density 
can aid in generating fallback. While the clouds that remain at 100 \msunpctwo{} are completely stripped except for the innermost clouds, only a fraction of the cooling clouds get accelerated onto a trajectory consistent with full stripping. The rest of the clouds fall back and remain in the galaxy disk, having become too dense to strip. 

\section{Importance of Each Mechanism and Comparisons to Observations}
\label{sec:importance-and-comparisons}

Having established the possible mechanisms for fallback in the previous section, we now discuss when during an RP event each mechanism can contribute to fallback. We also discuss which mechanisms are responsible for more inclined winds causing higher levels of fallback. Lastly, we compare our findings with instances of ISM fallback observed in real RPS galaxies.

\subsection{Importance of Each Mechanism}
\label{sec:mechanism-importance}

While we are limited in our claims of which processes might dominate for a given galaxy, we can infer when some processes are more effective at contributing to fallback when considering the different stages of a ram pressure event. 
Furthermore, it is likely that multiple mechanisms can contribute at the same time, and may compliment each other.
The evolution of an infalling disk galaxy's ISM can be divided into two approximate epochs, with the first when the gas disk has considerable rotational motion, and the second when the gas disk is more truncated and lacks prominent rotation. 
As an example, the 78DEG run loses most of its disk rotation before pericentric passage. Following this, it enters a later stage where fallback still occurs but the tail is now thin and aligned with the galaxy center , instead of being pushed to one side like how it is observed in Figure \ref{fig:stfb}.

Naturally, spatially offset rotation (Section \ref{sec:spatially-offset-rotation}) can only contribute when the disk has prominent rotational motion.
Reductions in ram pressure strength downstream from the remaining disk due to shadowing (Section \ref{sec:shadowing}) also occurs during this epoch, but can also contribute during the later-stage mode. This is especially true if the galaxy is moving supersonically and has a bow shock, which is the case for our simulations.
Spatially-offset rotation and shadowing together help explain why a more inclined wind generates more fallback overall, since a highly inclined wind generates more asymmetrical effects through torques on the ISM gas due to the rotation of the disk.
Furthermore, only in an inclined wind will tail material orbit into the shadowed region (i.e. in a face-on wind, the wind never stops ``seeing" the disk material).
It also explains why fallback is predominantly occurring on the trailing side of the galaxy rotating into the wind (Q4), since the orbits of these perturbed but un-stripped clouds are going to be moving back towards the disk in this quadrant.
Global reductions as the galaxy leaves pericenter (Section \ref{sec:post-peak}) may aid in generating fallback in later stages when the remaining gas is truncated and lacks strong rotation. Indeed, a global drop in wind strength could cause fallback of tail gas back into a completely stripped disk.
We note that only at late times in the 78DEG simulation is there a net infall of all ISM gas rather than only cold (dense) gas.
With decreasing ram pressure, lower density gas should be able to fall back to the disk, which may be a factor contributing to the lower cold gas fraction of fallback gas seen at later times.

Lastly, cloud cooling and compression (Section \ref{sec:cloud-compression}) can contribute at any time during an RP event. Since the mass growth of a cloud can be expressed as $\dot{M} \propto v_\text{rel} / \chi^{0.5}$, where $\chi = \rho_\text{wind} / \rho_\text{cl}$ is the density contrast between the wind and the cloud \citep[See][]{fieldingStructureMultiphaseGalactic2022}, clouds may entrain more mass as both the relative velocity and the density of the ICM increase near pericenter.
We note that cloud could also be heated and destroyed, which would result in less fallback.
While a detailed study of cloud evolution in different cluster and tail environments is beyond the scope of this work, we can conclude that if clouds increase their surface density they will be more likely to fallback onto the disk.

\subsection{Comparison to Known Observations of Fallback}
\label{sec:comparisons-to-observations}

Lastly, we consider the implications of our findings on the literature examples for gas fallback in two Coma cluster galaxies: NGC 4858 \citep{souchereauALMAJELLYHighResolution2025} and NGC 4921 \citep{cramerMolecularGasFilaments2021}. 
Both galaxies have CO(2-1) observations taken by the ALMA telescope.

\textbf{NGC 4858} is barred spiral jellyfish galaxy in the Coma cluster with an estimated stellar mass of $M_* = 4.9\times10^9$\msun   
\citep{molnarWesterborkComaSurvey2022}. It has a projected velocity relative to the cluster mean of $> 2400$ \kms. It has a strong molecular gas asymmetry with two prominent tail arms located in Q4. It is currently experiencing strong ram pressure stripping, and is likely pre-peak in its evolution.
Analysis of the non-circular motions of the CO(2-1) gas reveal a strong feature at the base of the inner tail moving opposite to the direction expected from ram pressure.
Based on the location of the feature, as well as the observed direction of the RPS tail, \cite{souchereauALMAJELLYHighResolution2025} determined that this gas must be falling back, as it is moving either radially inward or vertically back towards the mid-plane.

Because NGC 4858 is still infalling towards pericenter, still has circular rotation in its disk, and its disk-wind angle was estimated to be highly inclined  ($\phi_\text{DW} = 75_{-27}^{+10}$ degrees), the fallback is likely being driven primarily by spatially offset rotation combined with a shadow.
While no observational evidence for a bow shock exists for NGC 4858, its extremely high velocity through the Coma cluster would suggest that it has one that may be reducing the wind strength downstream.

\textbf{NGC 4921} is a massive ($M_* \approx 9\times10^{10}$ \msun{} using the mass-to-light relation in \cite{cluverGALAXYMASSASSEMBLY2014} and magnitudes from the ALLWISE catalog \citep{wrightWidefieldInfraredSurvey2010}) spiral galaxy in the Coma cluster with a projected velocity relative to the cluster center of $-1500$ \kms.
\citep{cramerMolecularGasFilaments2021} determined that clouds were falling back vertically towards the disk mid-plane in NGC 4921.  However, these fallback clouds are located in Q1, or the leading side of the galaxy rotating into the wind.
This may imply that the clouds are at a later stage of their orbit than the fallback detected in NGC 4858, having orbited all the way back to the leading side of the galaxy.
The galaxy still has clear rotational motion which would imply, as for NGC 4858, that rotation and shadowing are important factors.
This is especially true as its high mass likely limits the ability of the ram pressure wind to easily strip the remaining cold gas, suggesting that the gas may be perturbed but not accelerated to the escape speed over one rotation.

\section{Summary and Conclusion}
\label{sec:conclusion}

In this work we present the results of a suite of high-resolution wind tunnel simulations run using the \enzo{} adaptive mesh refinement code. 
We ran three simulations, changing only the angle in which the wind was injected ($45\degree$, $60\degree$, and $78\degree$). The velocity and density of the wind was tuned to match a realistic orbit of a satellite galaxy through the Coma cluster, with a pericentric passage (peak ram pressure) approximately 600 Myr after the wind arrives at the disk. 
Using these simulations, we develop a physically-motivated prescription to measure and quantify gas that has been pushed out of the ISM disk by ram pressure and subsequently falls back in, or ``fallback gas" (Section \ref{sec:identifying-fallback}). We measure the quantity of fallback across our simulation volumes using the \dtail-\vtail{} criteria, and also measure the mass flow rates over time in spherical shells of different radii (Section \ref{sec:quantifying-fallback}). To further examine the physical mechanisms contributing to gas fallback, we conduct a set of idealized dynamical simulations, tracing gas clouds under RPS as test particles. 

Based on our \enzo{} simulations and idealized dynamical modeling, we present the following conclusions:

\begin{enumerate}
    \item \textbf{Satellites experiencing an inclined wind take longer to strip and have more fallback.} As expected from other wind-tunnel simulations that cover a range of inclination angles, the more inclined (closer to edge-on) the ram pressure wind, the longer it takes to remove the ISM of a satellite \citep[Figure \ref{fig:disk-gas-mass-evol}; e.g.][]{roedigerRamPressureStripping2006, jachymRamPressureStripping2009}. 
    We have shown here that more inclined ram pressure winds tend to generate elevated quantities of gas falling back into the gas disk, where amount of fallback for the 45DEG run peaks at $1.8\times10^8$\msun, but the 78DEG run peaks at $3.2\times10^8$\msun.
    
    \item \textbf{Fallback gas is colder on average.} Cold gas generally makes up 50\% of the total fallback gas initially, and the proportion increases with time. The cold gas fraction of fallback is $\sim2-20$ times higher than the cold gas fraction of all gas in the tail (Figure \ref{fig:stripped-fallback-evol}). This agrees with the reduced capacity for ram pressure to strip higher-density gas, as ram pressure acceleration is inversely proportional to surface density.

    \item \textbf{Fallback is not a short-lived phenomenon.} Except for at the earliest and latest times, fallback can be detected throughout the duration of an RPS event (Figures \ref{fig:stripped-fallback-evol} and \ref{fig:mass-fluxes}). Most fallback, however, appears to occur before the galaxy reaches pericenter.

    \item \textbf{Fallback tends to be on the side of the galaxy rotating into the wind (Quadrant 4), and closer to the disk.} Fallback is concentrated in Q4 especially at early times, but becoming less concentrated over time (Figure \ref{fig:stripped-fallback-evol}). In our simulations, the majority of inward gas flow is occurring within 20 kpc of the galaxy center, regardless of disk-wind angle (Figure \ref{fig:mass-fluxes}). This is consistent with density projections tagging fallback in our snapshots, which show that most fallback appears in clouds or arm-like structures close to the remaining gas disk. Beyond 20kpc, fallback is only detected in compact, dense clouds.

    \item \textbf{Offset-Rotation is sufficient to generate fallback.} Even in the absence of other possible mechanism, an inclined wind and galaxy rotation can generate fallback (Figure \ref{fig:inward-outward-orbits}). Offset rotation will also create fallback in Q4, as the offset orbits are returning to the disk in this quadrant.

    \item \textbf{Fallback can be generated by gas moving into areas of lower local ram pressure.} Whether through hydrodynamical ``shadowing" as the wind runs into the gas disk, or through a bow shock, galaxies can generate local downstream reductions in ram pressure (Figure \ref{fig:shadow-profiles-example}). For our simulations, where our galaxy is traveling at $\mathcal{M}\sim3-5$, the shadowed region can reduce ram pressure strength almost entirely at the line of symmetry along the tail. The shadow can persist even as the ISM disk becomes very small and compact, and only dissipates when the galaxy becomes totally stripped.

    \item \textbf{A rotating disk is not required for fallback.} Shadowing downstream, as well as a drop-off of ram pressure strength after the pericentric passage, can contribute to later-stage fallback even when the galaxy lacks a rotating disk (Figures \ref{fig:inward-outward-orbits-shadowing} and \ref{fig:orbits-post-peak}). Furthermore, even modest increases in individual cloud surface densities can drastically alter their orbits, and can have a considerable impact on the quantity of fallback (Figure \ref{fig:orbits-cloud-cooling}).

\end{enumerate}

Based on our findings, the fallback in NGC 4858, which has the strongest detection for fallback observed in an RPS galaxy to date, is found in the most common region for fallback in a highly-inclined disk (i.e. Q4, or the trailing side of the galaxy on the side rotating into the wind).

We also predict that highly resolved searches could find fallback in nearly all galaxies undergoing ram pressure stripping, but because it will be spatially coincident (in 2D projection) with out-flowing gas, high spectral resolution will be required to identify and separate multiple velocity components. 
Also, galaxies post-pericenter in their orbits should have more fallback if they have not been fully stripped. Therefore, we recommend searching for jellyfish galaxies with tails pointing towards the cluster center. 
Such samples exist for future explorations, such as \cite{robertsLoTSSJellyfishGalaxies2021a} and \cite{salinasConstrainingDurationRam2024a}.

In future work, we will study how much of the fallback gas consists of cold gas created by in-situ cooling in the RPS tail.
We will also look more carefully at how this fallback gas affects the disk spatial and velocity structure. While we find that in these simulations the mass of gas falling back at any time is low, it could have an outsize impact on a heavily-stripped disk.
Indeed, if we require the surviving disks of galaxies to be unperturbed in order to unequivocally assign ram pressure stripping, even a small amount of fallback could confuse our understanding of how ram pressure affects galaxies.


\begin{acknowledgments}

We thank the anonymous referee for their thoughtful comments and suggestions which helped strengthen the paper.
The authors would also like to thank Greg Bryan, Isabel Medlock, and Frank Van den Bosch for helpful discussion and advice.

HS acknowledges the support from the NRAO Student Observing Support (SOS) grant SOSPADA-031.  HS also acknowledges support from from the CCA Pre-doctoral Program, during which this work was initiated.

The simulations used in this work were run and analyzed on facilities supported by the Scientific Computing Core at the Flatiron Institute, a division of the Simons Foundation.
This publication makes use of data products from the Two Micron All Sky Survey, which is a joint project of the University of Massachusetts and the Infrared Processing and Analysis Center/California Institute of Technology, funded by the National Aeronautics and Space Administration and the National Science Foundation.

\end{acknowledgments}

\software{\textsc{Enzo} \citep{bryanENZOAdaptiveMesh2014}, yt \citep{turkYtMulticodeAnalysis2011}, Astropy \citep{astropycollaborationAstropyCommunityPython2013, astropycollaborationAstropyProjectBuilding2018, astropycollaborationAstropyProjectSustaining2022}}

\appendix

\section{Effects of Simulation Resolution}
\label{app:caveats-resolution}

\begin{figure*}
    \includegraphics[width=\textwidth]{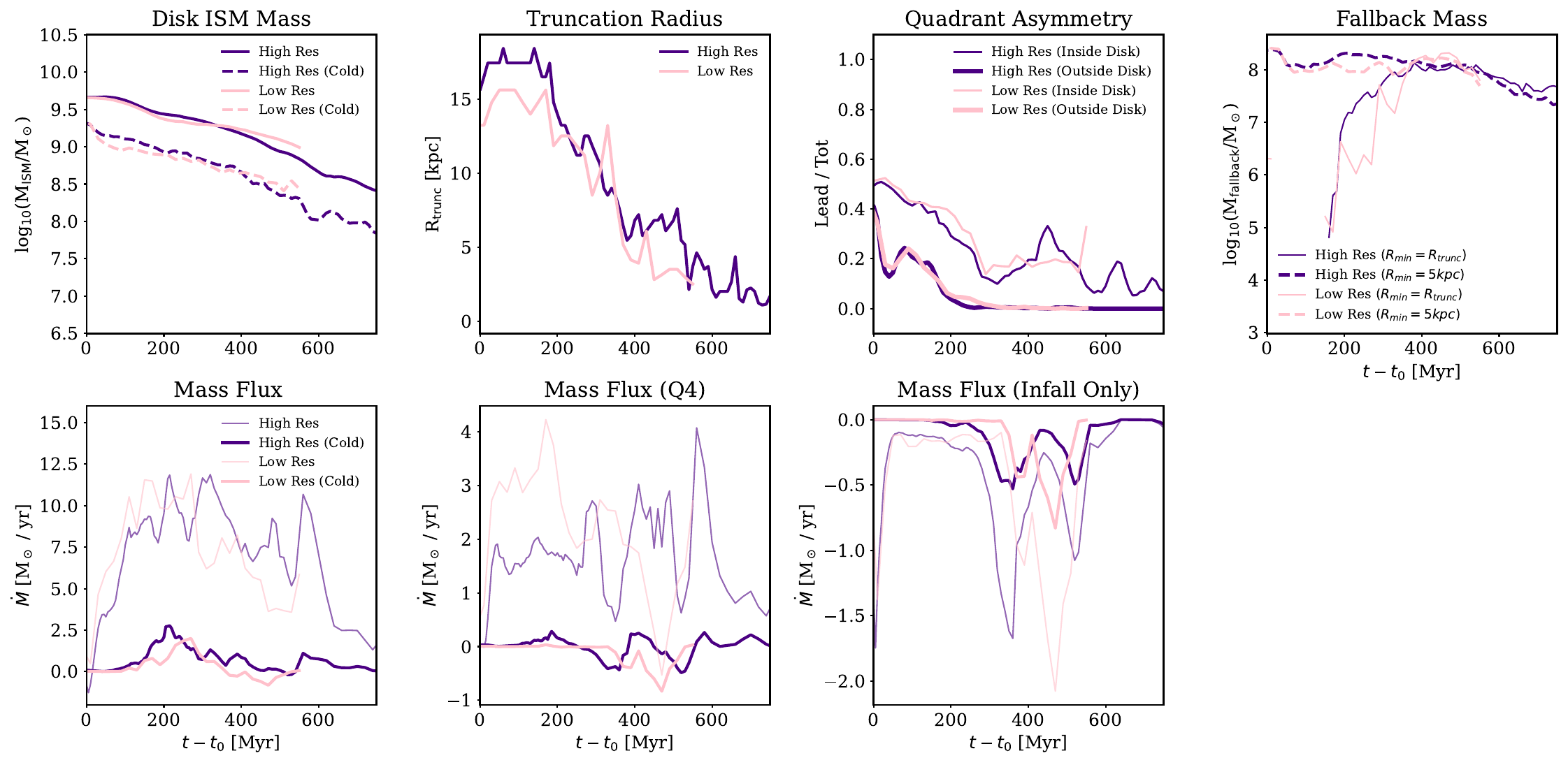}
    \caption{Comparison of key results between the high resolution 78DEG run (5 refinement levels, minimum cell size of 40 pc) and a lower resolution run (3 refinement levels, minimum cell size of 160 pc). In the top row, from left to right, we compare disk gas mass and truncation radius ($R_\text{trunc}$) (see Figure \ref{fig:disk-gas-mass-evol}), quadrant asymmetry (see Figure \ref{fig:quadrant-mass-evol_cold}), and fallback gas mass (see Figure \ref{fig:stripped-fallback-evol}). In the second row we compare mass fluxes over the entire spherical shell, the Q4 quarter sphere, and infalling gas only (see Figure \ref{fig:mass-fluxes}).}
    \label{app:resolutiontest_summary}
\end{figure*}

Here we compare several measures between our fiducial high resolution (smallest cell size = 40 pc, in purple) and low resolution run (smallest cell size = 160 pc, in pink, summarized in Figure \ref{app:resolutiontest_summary}.

In terms of total gas mass, we see a good agreement between the high and low resolution runs for both ISM and cold ISM. 
The distribution of cold gas is slightly different, however, as measured by the truncation radius. We see that the low resolution run is more compact than the high resolution run by a few kpc at early times, and then perhaps again at late times.
However, the overall trend between the simulations is quite similar.

The relationships between the leading and trailing side quadrants agree with each other (``Quadrant Asymmetry'' panel), indicating that larger-scale behavior is consistent.
The fallback mass values show minor differences when using $d_{\rm proj, min} = R_{\rm trunc}$ as our distance cut for fallback. 
There is less fallback in the lower resolution run at earlier times, becoming slightly elevated compared to the high resolution run at later times.
However, it is evident that the qualitative behavior of the time evolution of fallback mass is consistent.

While the mass fluxes differ between the low and high resolution runs, our main results hold (three bottom panels). 
Overall net fallback is only seen in the cold material, and Q4 seems to be the site of most of the infall. 
When measuring the amount of infall, the peaks for infall differ, where the first peak of infall is 350 Myr for the high resolution run, where it is $\sim 475$ Myr for the low resolution run. 
This is consistent with the heightened fallback in the low resolution run measured using the \dtail-\vtail{} method around this time. However, as shown in the ``Fallback Mass'' panel, the overall amount of fallback is quite similar.  

Altogether, these panels show that while changing resolution may quantitatively change our individual measurements of the spatial and velocity distribution of the gas, our qualitative results remain unchanged.

\section{\dtail-\vtail{} Time Series}
\label{app:projection-timeseries}

\begin{figure*}
    \includegraphics[width=\textwidth]{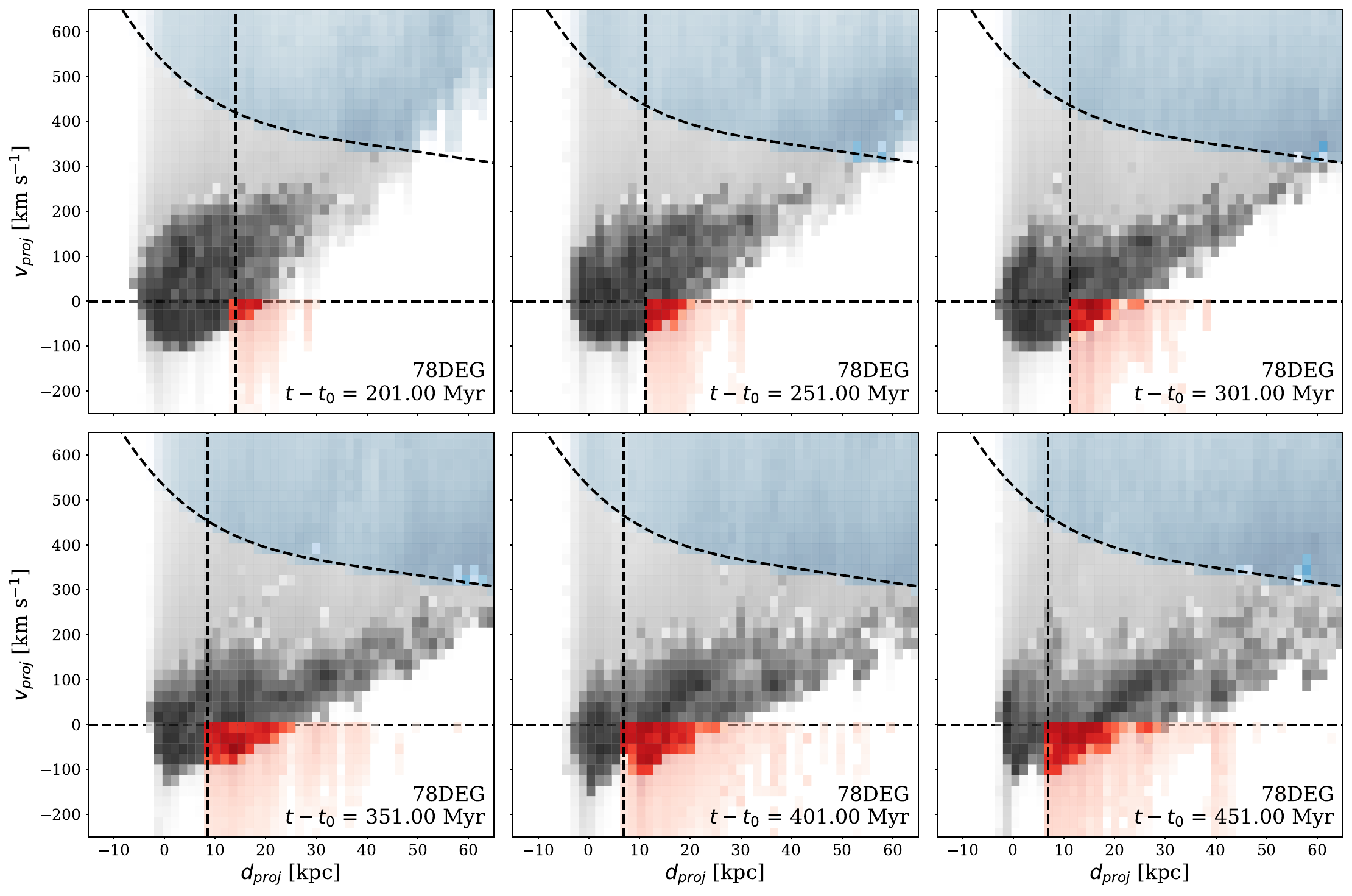}
    \caption{The same as the left panel in Figure \ref{fig:stfb}, but over multiple snapshots of the 78DEG run to show evolution in \dtail-\vtail{} phase space.}
    \label{app:dproj-vproj-timeseries}
\end{figure*}

In Figure \ref{app:dproj-vproj-timeseries} we show the evolution of gas in the \dtail-\vtail{} phase space at time steps of 50 Myr. 
Although we cannot follow individual gas cells through our simulations, we can see in these panels that the general behavior of un-stripped tail gas is to be stretched out to higher \dtail{} values. 
The extent of cold fallback gas steadily increases from $\sim15$ kpc at $t-t_0 = 201$ Myr to $\sim30$ kpc at $t-t_0 = 451$ Myr, but remains concentrated close to the boundary of the disk.
This steady increase in the extent of fallback gas, even while the ram pressure strength is increasing, indicates that gas is following the trajectories illustrated in Figure \ref{fig:stfb-schematic}.

\bibliography{main}{}
\bibliographystyle{aasjournal}

\end{document}